\documentclass[12pt]{article}
\pdfoutput=1
\usepackage[nosort]{cite}

\usepackage{xcolor}

\usepackage{epsfig}
\usepackage{amsfonts}
\usepackage{amscd}
\usepackage{latexsym}
\usepackage{amsmath,amssymb}
\usepackage{verbatim}
\usepackage{setspace}
\usepackage{color}
\usepackage{fancyhdr}
\usepackage{cite}
\usepackage{hyperref}
\usepackage{tikz-cd}
\usepackage{slashed}
\usetikzlibrary{calc}
\usetikzlibrary{topaths}
\usetikzlibrary{decorations}
\usetikzlibrary{decorations.pathmorphing}

\usepackage{draft}
\usepackage{hyperref}
\usepackage{graphicx,color,subfig}
\usepackage{cite}
\usepackage{mciteplus}
\usepackage{skak}

\numberwithin{equation}{section}

\newcommand{\lxeff}{\ell_x^{\textrm{eff}}}
\newcommand{\lyeff}{\ell_y^{\textrm{eff}}}
\newcommand{\deltaP}{\delta^{\textrm{P}}}
\newcommand{\ThetaP}{\Theta^{\textrm{P}}}
\newcommand{\Lxeff}{L_x^{\textrm{eff}}}
\newcommand{\Lyeff}{L_y^{\textrm{eff}}}

\newcommand{\bW}{\textbf{W}}
\newcommand{\bP}{\textbf{P}}

\begin{document}

\begin{titlepage}

\begin{center}

\hfill \\
\hfill \\
\vskip 1cm

\title{Fractons with Twisted Boundary Conditions and Their Symmetries}

\author{Tom Rudelius$^{1, 2}$, Nathan Seiberg$^{2}$, and Shu-Heng Shao$^{2}$}

\textit{\centerline{${}^{1}$Physics Department, University of California, Berkeley CA, USA}
\centerline{${}^{2}$School of Natural Sciences, Institute for Advanced Study, Princeton NJ, USA}
}

\end{center}

\vspace{2.0cm}

\begin{abstract}
\noindent
We study several exotic systems, including the X-cube model, on a flat three-torus with a twist in the $xy$-plane.  The  ground state degeneracy turns out to be a sensitive function of various geometrical parameters.  Starting from a lattice, depending on how we take the continuum limit, we find different values of the ground state degeneracy.  Yet, there is a natural continuum limit with a well-defined (though infinite) value of that degeneracy. We also uncover a surprising global symmetry in $2+1$ and $3+1$ dimensional systems. It originates from the underlying subsystem symmetry, but the way it is realized depends on the twist. In particular, in a preferred coordinate frame, the modular parameter of the twisted two-torus $\tau = \tau_1 + i \tau_2$ has rational $\tau_1 = k / m$.  Then, in systems based on $U(1)\times U(1)$ subsystem symmetries, such as momentum and winding symmetries or electric and magnetic symmetries, the new symmetry is a projectively realized $\bZ_m\times \bZ_m$, which leads to an $m$-fold ground state degeneracy. In systems based on $\bZ_N$  symmetries, like the X-cube model, each of these two $\bZ_m$ factors is replaced by $\bZ_{\gcd(N,m)}$.

\end{abstract}

\vfill

\end{titlepage}

\tableofcontents

\section{Introduction}\label{se}

The exciting, growing field of fracton phases of matter  started with the discovery of two peculiar models  \cite{PhysRevLett.94.040402,PhysRevA.83.042330}.  They have stimulated a lot work, which has uncovered additional models of fractons and has led to deeper insights.   This subject is reviewed nicely in \cite{Nandkishore:2018sel,Pretko:2020cko}.  These reviews include many references to other interesting papers.

These models are not rotationally invariant, and the Hamiltonian depends on preferred directions, which we will denote by $(x,y,z)$.  They are typically formulated on a lattice with $L_x$, $L_y$, and $L_z$ sites in these directions with periodic boundary conditions. Then, the number of ground states depends on these three integers.  Models based on $\mathbb{Z}_N$ spins typically have ground state degeneracy
\ie
\text{GSD} =N^{Q(L_x,L_y,L_z)}~,
\fe
but, as we will see, other functional forms are also possible.
A characteristic example, which will also be studied below, is that of the X-cube \cite{Vijay:2016phm}, or more generally, its $\mathbb{Z}_N$ version, where the entropy $Q$ is given by
\ie\label{QXcube}
Q(L_x,L_y,L_z)=2(L_x+L_y+L_z)-3~.
\fe
This expression is peculiar for two reasons.  First, even though the system is gapped, the number of ground states diverges as the system size goes to infinity, i.e., in the limit $L_x,L_y,L_z \to \infty$.  Second, the expression \eqref{QXcube} is not extensive.  It is sub-extensive; it grows linearly with the size of the system.  Other examples, including the original Haah code \cite{PhysRevA.83.042330}, exhibit an even more bizarre $Q(L_x,L_y,L_z)$, which is not even monotonic in the three sizes.

The existence of these models raises many interesting and deep questions.  One of them is how to formulate a continuum quantum field theory description of them.  Early work on the subject appeared in  \cite{Slagle:2017wrc,Slagle:2018swq, Slagle:2020ugk}.  Here we will follow the systematic approach of \cite{Seiberg:2019vrp,paper1,paper2,paper3,Gorantla:2020xap,Gorantla:2020jpy}.

The original models were formulated on a flat right-angled torus aligned with the preferred directions $(x,y,z)$.  This immediately raises the question how to formulate these models on more complicated manifolds.  An important idea in this direction is to place the system on a foliated space \cite{Slagle:2018wyl,Shirley:2017suz,Shirley:2019wdf,Shirley:2019xnp,Shirley:2018vtc,Slagle:2018kqf,Slagle:2018swq,
Shirley:2020gfi,Slagle:2020ugk}.  The foliation then determines the alignment of the preferred coordinates $(x,y,z)$.

Our goal here is to place such a system on a slightly nontrivial space such that the analysis is still straightforward.  We will keep the torus flat, but will allow it to be slanted -- not right-angled.
We will also allow a twist of the torus relative to the preferred $(x,y,z)$ coordinate system.
The local interaction is still invariant under the appropriate subgroup of the rotation group, but the global boundary conditions do not respect this rotation symmetry.

\subsection{The twisted torus}

Specifically, we will study the system with twisted boundary conditions.  On the lattice, we label the sites by integers $(\hat x, \hat y, \hat z)$ and impose the identifications
\ie\label{generalrI}
(\hat x, \hat y, \hat z) \sim (\hat x+L_x^u, \hat y+L_y^u, \hat z+L_z^u)\sim (\hat x+L_x^v, \hat y+L_y^v, \hat z+L_z^v)\sim (\hat x+L_x^w, \hat y+L_y^w, \hat z+L_z^w)~.
\fe

Related problems were studied in \cite{Slagle:2018wyl,Shirley:2017suz,Dua:2019ucj,1821601,manoj2020screw,Meng}.   Although our approach is different, some of the issues we will address have counterparts in these papers.

Actually, for simplicity, we will limit ourselves to nontrivial twists only in two of the directions, i.e.,
\ie\label{rotationI}
(\hat x, \hat y, \hat z) \sim (\hat x+L_x^u, \hat y+L_y^u, \hat z)\sim (\hat x+L_x^v, \hat y+L_y^v, \hat z)\sim (\hat x, \hat y, \hat z+L_z)~.
\fe
We will refer to the closed cycles associated with these identifications as the $U$, $V$, and $Z$ cycles, respectively.

There is a lot of freedom in choosing the generators of the identifications. We will take all the integer coefficients $L_i^r$ to be non-negative.

Without loss of generality, we can also align the $U$ cycle with the $x$ direction -- the $X$ cycle.  Then, in order to have a complete basis, we need $V$ to be dual to $X$, the $\tilde X$ cycle.   In this case $L_y^u=0$, and some of our expressions below simplify. Alternatively, we can align the $V$ cycle with the $y$ direction -- the $Y$ cycle.  In this case, we need $U$ to be dual to $Y$, the $\tilde Y$ cycle.  It is important to note that in general, the $X$ and $Y$ cycles do not generate all the cycles, and therefore they cannot be used as a complete basis.  This fact will have interesting consequences.

We have analyzed all the models in \cite{paper1,paper2,paper3} on such a torus.  Some of these models are gapless.  Their states with generic momenta have a peculiar dispersion relation, but other than that, they are quite standard.  As these modes reflect local physics, the effect of the twisted boundary conditions on them is quite trivial.  These gapless theories also have strange states at non-generic momenta -- specifically, states where two of the momenta $p_x$, $p_y$, $ p_z$ vanish.   Some peculiarities of these modes were discussed in \cite{paper1,paper2}.\footnote{As emphasized in \cite{paper1,paper2}, some of the detailed features of the charged states in the gapless models depend on higher-derivative terms that go beyond the leading order terms in the continuum Lagrangian.  This subtlety is not present in the gapped models and does not affect the peculiarities we will discuss below.}

Here we will focus on the consequences of the twisted boundary conditions and will find that the system has  states that realize  the underlying subsystem symmetry in a surprising way.  In some of the non-gauge systems, some momentum and winding symmetries do not commute.  In some of the gauge theories, some electric and magnetic symmetries do not commute.  These effects are reminiscent of effects found in \cite{Freed:2006ya, Freed:2006yc} and discussed further in \cite{Kapustin:2014gua,Gaiotto:2014kfa}.

The gapped models are particularly interesting, and we will follow and extend their analysis in \cite{paper1,paper3}.  The twisted boundary conditions change the ground state degeneracy and the surprising realization of the subsystem symmetry in some gapless models has counterparts in the gapped systems.

Analyzing the X-cube model along the lines of \cite{paper3}, we will show that in this case \eqref{QXcube} is replaced by
\ie\label{QXcuberI}
\text{GSD}=N^{2(\Lxeff+\Lyeff+L_z)-3}\text{gcd}(N,M) \,,
\fe
where
\ie\label{QXcuberId}
&\Lxeff =\text{gcd}(L_x^u, L_x^v) \,,\\
&\Lyeff = \text{gcd}(L_y^u, L_y^v)\,,\\
&M={L_x^uL_y^v-L_x^vL_y^u \over \Lxeff\Lyeff}.
\fe
As stated above, without loss of generality we can take $L^u_y=0$, and then these expressions simplify:
\ie\label{QXcuberIs}
&\Lxeff =\text{gcd}(L_x^u, L_x^v) \,,\\
&\Lyeff = L_y^v\,,\\
&M={L_x^u\over \text{gcd}(L_x^u, L_x^v)}.
\fe
A special case of this expression was found in \cite{Meng}.

The ground state degeneracy \eqref{QXcuberI} has several interesting features.
\begin{itemize}
\item As in the untwisted model, the ground state degeneracy \eqref{QXcube} depends on the number of sites in the lattice.  As we rescale the lattice data to infinity $L_i^r\to \infty$ with fixed ratios, the number of ground states diverges in a sub-extensive manner.
\item Relative to the untwisted model, the number of ground states \eqref{QXcuberI} depends on more lattice data $L_i^r$.  Small changes in these integers can make a large effect on the number of ground states.  In fact, the ground state degeneracy does not change monotonically with this data.  These facts are reminiscent of the dependence of the ground state degeneracy on the number of sites in the Haah code \cite{PhysRevA.83.042330}.
\item As in the Haah code \cite{PhysRevA.83.042330}, the previous point makes it clear that the model does not have an unambiguous continuum limit.  Unlike the original untwisted model, where the logarithm of the ground state degeneracy diverges linearly in the size, but is otherwise well-defined, here different ways of taking the continuum limit lead to different answers.
\item The exponential dependence of the ground state degeneracy \eqref{QXcuberI} on $L_i^r$ has a natural interpretation in the layer constructions of these models \cite{Ma:2017aog,Vijaylayer}.  The model is constructed out of $L_z$ layers in the $xy$-plane, $\Lxeff$ layers in the $yz$-plane, and $\Lyeff$ layers in the $xz$-plane.  The exponential part of the degeneracy is then as in the untwisted model with the same number of layers.  The connection to the layers construction was discussed in a special case in \cite{1821601}.
See also the general discussions in \cite{Shirley:2017suz}, which advocates the use of foliated manifolds.
\item In addition to the exponential behavior in \eqref{QXcuberI}, there is also a factor of $\text{gcd}(N,M)$.  It reflects an interesting symmetry group, which is a central extension of $\mathbb{Z}_{\text{gcd}(N,M)}\times \mathbb{Z}_{\text{gcd}(N,M)}$. We will discuss it in detail below.
\end{itemize}

These peculiarities of \eqref{QXcuberI} follow from properties of the charges of the subsystem global symmetry (or equivalently,  the logical operators) of the system.
Some of these charges are associated with closed lines along $x$, or $y$, or $z$.
Because of the twisted boundary conditions \eqref{generalrI}, \eqref{rotationI}, these lines wrap the torus an integer number of times.  Consequently, the number of distinct charges depends sensitively on $L_i^r$.  The ground state degeneracy follows from the number of such charges.  This sensitivity leads to the peculiarities of the ground state degeneracy mentioned above.  This fact is reminiscent of the way the ground state degeneracy arises in the Haah code.

Let us comment on the continuum limit in more detail.  The continuum limit is taken by introducing a lattice spacing $a$ and taking $L_i^r \to \infty$ with fixed
\ie\label{ellirc}
\ell_i^r = \lim_{a\to 0} aL_i^r ~.
\fe
The fact that $L_i^\textrm{eff}$ can diverge in this limit and can lead to infinite $Q$ is common in these models.  The important point here is that the limits $\lim_{a\to 0} L_i^\textrm{eff}$ and $\lim_{a\to 0} a L_i^\textrm{eff}$ can depend on the way we take the continuum limit.  This means that different sequences of lattice models, all approaching $L_i^r \to \infty$ with the same continuum values \eqref{ellirc}, can have different ground state degeneracies.

This might lead us to question to what extent the continuum Lagrangian describes the physics of such a system.  The system must be regularized, and the limit as the regularization is removed can lead to an infinite ground state degeneracy that depends sensitively on the regularization.  However, there is a natural way to regularize the continuum system such that the answer is unambiguous.  In particular, we let integers $L_i^r$ go to infinity in fixed ratios.  More explicitly, starting with the continuum quantities $\ell_i^r$, we introduce a lattice spacing $a$ with lattice integers $L_i^r$ such that $aL_i^r =\ell_i^r$.  (This is possible only when the ratios of $\ell_i^r$ are rational.)

Taking this natural limit, we find the continuum limits of \eqref{QXcuberIs}:
\ie\label{QXcuberId}
&\lxeff =\lim_{a\to 0}a\Lxeff \,,\\
&\lyeff =\lim_{a\to 0}a\Lyeff \,,\\
&m =\lim_{a\to 0}M\,.
\fe
This means that in the continuum, the torus in the $xy$-plane is subject to the identifications
\ie\label{torusiden}
&(x,y)\sim (x+m\lxeff,y) \sim (x+k\lxeff, y+\lyeff) \qquad ,\\
& m,k\in \bZ\qquad,\qquad\text{gcd}(m,k)=1 \,.
\fe
The real part of the modular parameter $\tau=\tau_1+i\tau_2$ for this torus is rational, i.e.,  $\tau_1={k\over m}$.

We would like to stress an important point about the integers $m$ and $k$.  From \eqref{torusiden}, they appear to be related to the geometry of the torus rather than its topology.  However, the integers $m$ and $k$ have a topological meaning.  As we will discuss below, they are associated with intersection numbers of preferred cycles on the torus.  One way to realize their topological nature is to replace the metric $ds^2=dx^2+dy^2$ in the $xy$ coordinate system with another flat metric.  Then $\tau$ will be different, but the intersection numbers will not change.

\subsection{A new, surprising symmetry}

The analysis of \cite{paper1,paper2,paper3} starts with the $2+1$-dimensional XY-plaquette model of \cite{PhysRevB.66.054526}.  We refer to its continuum limit as the $\phi$-theory.  Its Lagrangian is
\ie\label{phiti}
{\cal L}= {\mu_0\over 2} (\partial_0\phi)^2 - {1\over 2\mu}(\partial_x\partial_y\phi)^2 \qquad , \qquad \phi \sim \phi + 2\pi \Big(n^x(x) +n^y(y)\Big) ~.
\fe
 with $n^x(x), n^y(y) \in \mathbb{Z}$. The two operators
\ie\label{phicop}
&J_0=\mu_0 \partial_0\phi\, ,\\
&J^{xy}= -{1\over \mu }\partial^x\partial^y \phi
\fe
form the Noether current of a momentum $U(1)$ subsystem symmetry with the conservation equation
\ie
\partial_0J_0=\partial_x\partial_y J^{xy}\,.
\fe
The conserved $U(1)$ charges are
\ie
&\oint dx J_0\,,~~~~~~\oint dyJ_0\,.
\fe
They are conserved also on the twisted torus.
The number of independent conserved charges is infinite, and we discretize them on a lattice.
It was $L_x+L_y-1$ on an untwisted torus, and reduces to $\Lxeff+\Lyeff-1$ on a twisted torus.

The same local operators (up to rescaling) \eqref{phicop} lead to a conserved current for a winding $U(1)$ subsystem symmetry
\ie\label{windingJ}
&J_0^{xy}={1\over 2\pi} \partial^x\partial^y\phi\,,\\
&J= {1\over 2\pi }\partial_0 \phi\,,\\
&\partial_0J_0^{xy}=\partial_x\partial_y J\,,
\fe
with the conserved $U(1)$ charges
\ie\label{windingci}
&\oint dx J_0^{xy}\,,~~~~~~\oint dy J_0^{xy}\,.
\fe
Again, their number is reduced by the twist from $L_x+L_y-1$ to $\Lxeff+\Lyeff-1$. As argued in \cite{paper1}, all the states that are charged under these symmetries acquire large energy, of order $1\over a$, in the continuum limit.  A conservative approach simply ignores them.

We will see that the theory on the twisted torus \eqref{torusiden}  has another symmetry constructed out of the same momentum and winding currents.  It is a clock and shift symmetry generated by two operators $U$ and $\tilde U$ satisfying
\ie\label{ZmZm}
U^m=\tilde U^m=1,\qquad U\tilde U=e^{2\pi i\over m} \tilde UU\,.
\fe
  This symmetry is a central extension of  $\bZ_m\times \bZ_m$.\footnote{
 More precisely, the operators of the theory are in linear representations of $\bZ_m\times \bZ_m$.  So strictly, this is the symmetry group of the system.  This symmetry is realized projectively on the Hilbert space.  This can be interpreted as an 't Hooft anomaly in the symmetry. }
  Here the first factor can be interpreted as a momentum symmetry and the second factor as a winding symmetry.  Surprisingly, these two symmetries do not commute.

One consequence of the clock and shift algebra \eqref{ZmZm} is that every state in the Hilbert is in an $m$-dimensional representation.  In particular, the system \eqref{phiti} on the twisted torus \eqref{torusiden} with $\tau_1={k\over m}$ has $m $ ground states!

The same conclusion is true for the $3+1$-dimensional version of this model, which was analyzed on the untwisted torus in \cite{paper2}.  This model is dual to a gauge theory, the $\hat A$-theory \cite{paper2}.  In the language of this gauge theory, the theory has electric and magnetic subsystem symmetries, and the central extension of $\bZ_m\times \bZ_m$ represents non-commutativity between electric and magnetic fluxes.

This situation is reminiscent of the analysis of ordinary $U(1)$ gauge theories in $3+1$ dimensions on a manifold with torsion cycles  \cite{Freed:2006ya, Freed:2006yc}.  A cycle $\gamma$ in space is torsion if $\gamma$ is not contractible, but $m\gamma$ is contractible, i.e., there is a surface $\Sigma$ such that $m\gamma=\partial\Sigma$.  Following \cite{Kapustin:2014gua,Gaiotto:2014kfa}, we can interpret  \cite{Freed:2006ya, Freed:2006yc} as follows.  The operator
\ie
U=e^{i\oint_\gamma A-{i\over m}\int_\Sigma dA}
\fe
satisfies $U^m=1$, but $U$ itself is nontrivial. Similarly, using the dual gauge field $\tilde A$, the operator
\ie
\tilde U=e^{i\oint_\gamma \tilde A-{i\over m}\int_\Sigma d\tilde A}\,
\fe
satisfies $\tilde U^m=1$.
The parts  of these operators associated with the surface $\Sigma$ are similar to the charges of the magnetic one-form symmetry and the electric one-form symmetry respectively.  However, since they include also the Wilson and the 't Hooft lines, they are charged under the electric and the magnetic one-form symmetries respectively.  As a result, $U$ and $\tilde U$ do not commute and obey \eqref{ZmZm}.

In the case of the X-cube model, the $U(1)$ subsystem symmetry of the gauge theory is replaced by a $\bZ_N$ subsystem symmetry.  In that case, this central extension of $\bZ_m\times \bZ_m$ is changed to a central extension of $\bZ_{\gcd(N,m)}\times \bZ_{\gcd(N,m)}$.  Its irreducible representation is $\gcd(N,m)$-dimensional.  This leads to a factor of $\gcd(N,m)$ in the ground state degeneracy and corresponds to the factor of $\gcd(N,M)$ in the lattice expression \eqref{QXcuberI}.

Below we will discuss this symmetry and its consequences in much more detail.

We end this subsection by pointing out that this relation to \cite{Freed:2006ya, Freed:2006yc,Kapustin:2014gua,Gaiotto:2014kfa} and the analysis in Appendix \ref{app:Uphi0} suggest that our discussion can be phrased in an appropriate version of differential cohomology. (See an introduction for physicists in \cite{Freed:2006ya, Freed:2006yc, Kapustin:2014gua,Gaiotto:2014kfa, Bauer:2004nh, Cordova:2019jnf}.)  We will not do it here.

\subsection{Outline}

In Section \ref{sec:GEOMETRY}, we will discuss the geometry of the foliated torus.  For simplicity, we will focus on a two-torus.  We will first analyze a continuous torus and then discuss its lattice version.

In Section \ref{sec:2+1winding}, we will place a classical, circle-valued field $\phi \sim \phi+2\pi$
on our twisted torus and will explore its winding configurations.  Here we will find the $\bZ_m$ winding charges we mentioned above. This will lead us to a discussion of the symmetries and the spectrum of the $2+1$-dimensional $\phi$-theory of \cite{paper1} on the twisted torus.

In Section \ref{sec:2+1d}, we will study a $2+1$-dimensional $\mathbb{Z}_N$ tensor gauge theory on the twisted torus. This model was analyzed on an untwisted torus in \cite{paper1}.  Starting with a lattice, this model is not robust under small deformations of the lattice system.  However, as discussed in \cite{paper1}, it makes sense as a continuum field theory.  We will study its two dual continuum presentations of \cite{paper1}.  We will analyze the ground state degeneracy and the spectrum of operators.  We will also comment on the bundles and transition functions of the $2+1$-dimensional $U(1)$ $A$-theory of \cite{paper1} and the $2+1$-dimensional $\bZ_N$ tensor gauge theory on the twisted torus.

Section \ref{sec:3+1winding} will analyze the winding configurations of a circle-valued field on the twisted three-torus.  This information will be important in Section \ref{31ZN}, where we will use the various dual continuum field theory descriptions in \cite{paper3}  to analyze the $3+1$-dimensional X-cube model (i.e., the $3+1$-dimensional $\mathbb{Z}_N$ tensor gauge theory) on our twisted torus \eqref{rotationI}.

We will present some more technical information in  appendices.
In Appendix \ref{app:2+1lattice}, we will analyze the $2+1$-dimensional $\bZ_N$ plaquette Ising model  in the broken phase.  In the continuum limit it becomes the $\bZ_N$ tensor gauge theory of Section \ref{sec:2+1d} \cite{paper1}.  We will compute the ground state degeneracy on a twisted torus and match it with the answer from the continuum treatment. This provides a further check of our answer.
In Appendix \ref{app:counting}, we will discuss the invariants of the transition functions for a circle-valued field $\phi$  in Section \ref{sec:2+1winding}.
Appendix \ref{app:Uphi0} will discuss additional operators that lead to the $\bZ_m\times \bZ_m$ symmetry  in the $\phi$-theory.
The analogous operators in the $2+1$-dimensional $\bZ_N$ theory will be subsequently analyzed in Appendix \ref{app:ZN}.
Finally, Appendix \ref{sec:hatwinding} will discuss the winding configurations of a circle-valued field $\hat\phi^{i(jk)}$ in the $\mathbf{2}$ of $S_4$.

\section{Geometry}\label{sec:GEOMETRY}

In this section, we focus on the geometry of a flat two-dimensional torus $T^2$ on which we are going to place our system.

\subsection{Continuum geometry}

\begin{figure}
\centering
\includegraphics[width=75mm]{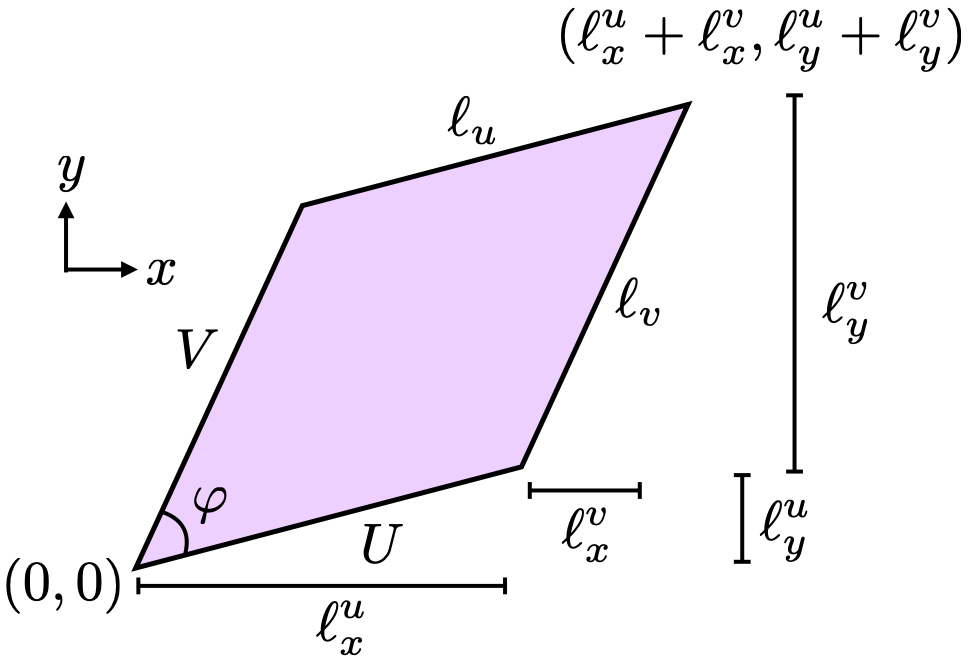}
\caption{The fundamental domain of the spatial torus.    }\label{fundamentaldomain}
\end{figure}

Our system is equipped with a preferred coordinate system $(x,y)$.  We place it on a torus by imposing identifications generated by
\ie\label{rotationxy}
( x,  y) \sim ( x+\ell_x^u,  y+\ell_y^u)\sim ( x+\ell_x^v,  y+\ell_y^v)~.
\fe
As in \eqref{rotationI}, we can take all  $\ell_i^r\ge 0$. These two identifications correspond to two cycles of the torus, which we denote by $U$ and $V$ respectively.
See Figure \ref{fundamentaldomain} for an illustration of this geometry.

The preferred coordinate system $(x,y)$ leads to a foliation of the torus.  It is given by the special lines of constant $x$ and constant $y$.  As we will see, the physical answers depend both on the parameters of the torus and on the choice of foliation.  For simplicity, we are going to limit ourselves to the case where these special lines wrap the torus a finite number of times.  Otherwise, some of the integers below are infinite.

\begin{itemize}
\item The $Y$ cycle of the torus is characterized by constant $x$.  It wraps the $V$ cycle $w_x^v$ times and it wraps the $U$ cycle $-w_x^u$ times.  $w_x^v$ and $w_x^u$ are non-negative integers satisfying $\gcd(w_x^v,w_x^u)=1$.  Using \eqref{rotationxy}, we have
    \ie
    \ell_x^u w_x^u=\ell_x^v w_x^v \,.
    \fe
\item The $X$ cycle of the torus is characterized by constant $y$. It wraps the $V$ cycle $-w_y^v$ times and it wrap the $U$ cycle $w_y^u$ times.  Again, $w_y^v$ and $w_y^u$ are non-negative integers satisfying $\gcd(w_y^v,w_y^u)=1$.   Using \eqref{rotationxy}, we have
    \ie
    \ell_y^u w_y^u=\ell_y^v w_y^v \,.
\fe
\end{itemize}
The condition that the $w_i^r$ must be finite integers amounts to the statement that $\ell_x^v/\ell_x^u$ and $\ell_y^v/\ell_y^u$ are rational.

More mathematically, consider the first homology group $\Gamma\equiv H_1(T^2,\bZ)\simeq \bZ^2$ of the torus with integer coefficients.
The lattice $\Gamma$ is generated by the $U$ and the $V$ cycles.
Their intersection  numbers are $\langle U,V\rangle = - \langle V,U \rangle=1$.  The $X$ and $Y$ cycles mentioned above are
\ie
&X=w_y^uU-w_y^v V\\
&Y=-w_x^uU+w_x^v V~.
\fe
The intersection between these two cycles is
\be\label{mdef}
m \equiv \langle X,Y\rangle=  \det\left(
\begin{array}{cc}
w_x^v& w_y^v \\
w_x^u & w_y^u \\
\end{array}
\right)  =  w_x^vw_y^u-w_y^vw_x^u  \, .
\ee
By exchanging the $U$ and the $V$ cycle, we can  take $m$ to be positive.

We will denote the sublattice generated by the $X$ and the $Y$ cycles by $\hat\Gamma$.  Using \eqref{mdef}, the index of this sublattice is $m$, i.e.,
\ie\label{Zmtor}
{\Gamma / \hat \Gamma} =\bZ_m\,.
\fe

When $m\ne 1$, $\hat \Gamma \subset\Gamma$ and the $X$ and $Y$ cycles are not a complete basis of $\Gamma$.  However, we can still choose a basis involving $X$.  It is related to the more generic $\{U,V\}$ basis by an $SL(2,\mathbb{Z})$ transformation
\ie\label{XtildeXbasis}
&\left(\begin{array}{c} X\\ \tilde X\end{array}\right)=\left(\begin{array}{cc} w_y^u & \, -w_y^v \\ - n_y^u&\, n_y^v\end{array}\right)\left(\begin{array}{c} U \\ V
\end{array}\right)\\
&w_y^u n_y^v-w_y^v n_y^u=1~,
\fe
where the condition on $n_y^v$ and $n_y^u$ can be satisfied because $\text{gcd}(w^u_y, w^v_y)=1$.  This defines the dual cycle $\tilde X = - n_y^u U + n_y^vV$.
The transformation $SL(2,\mathbb{Z})$ \eqref{XtildeXbasis} guarantees that the cycles $X$ and $\tilde X$ generate the entire lattice $\Gamma$, and their intersection is
\ie
\langle  X,\tilde X\rangle = 1~.
\fe
The cycle $\tilde X$ can be redefined further by adding to it an arbitrary integer multiple of $X$.

When $m\neq 1$, while $\tilde X$ is not an element of $\hat\Gamma$, $m\tilde X$ is. More explicitly,
\ie\label{hatY}
m\tilde X =  Y+ \langle \tilde X,Y\rangle  X \in \hat\Gamma\,.
\fe

The cycle $\tilde X$ can be taken to be the generator of $\bZ_m$ in \eqref{Zmtor}.  Intuitively, if we mod out by the cycles generated by $X$ and $Y$, we can think of $\tilde X$ as a torsion cycle.  This fact will have important consequences below.

\begin{figure}
\centering
~~~~~~\includegraphics[width=110mm]{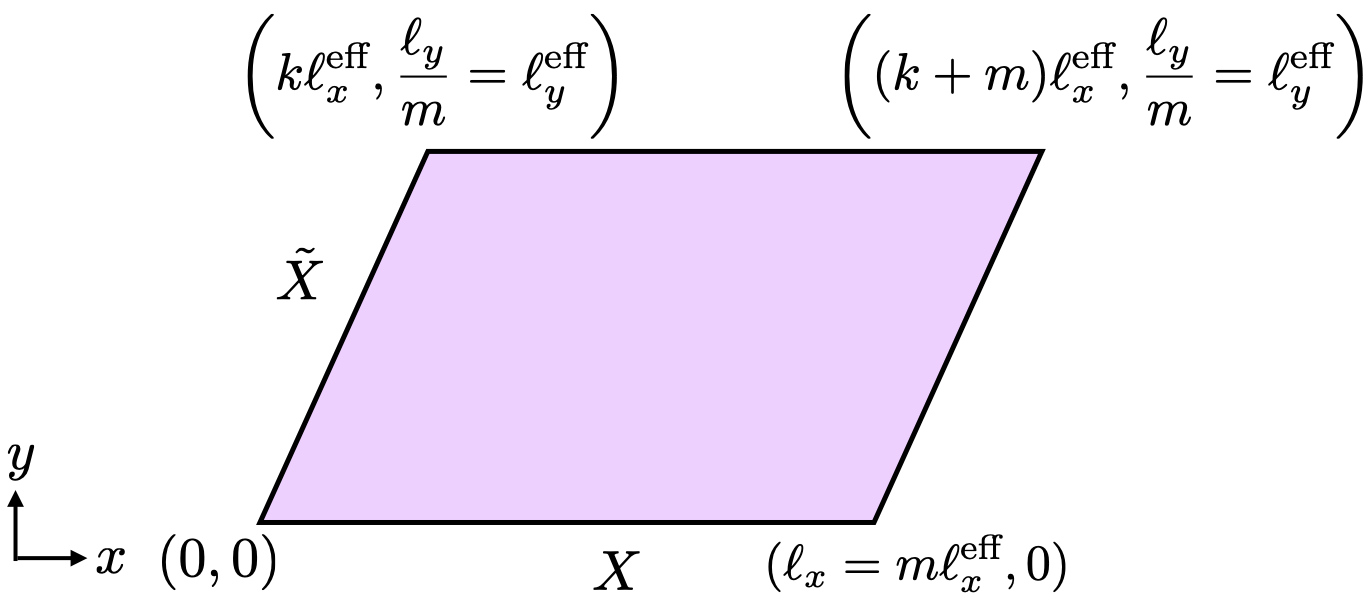}
\caption{The $X$ and $\tilde X$ cycles.  }\label{XXtcycles}
\end{figure}

Similarly, we can define the dual of the $Y$ cycle:
\ie\label{dualcycle}
&\tilde Y = -  n_x^u U + n_x^vV~,\\
&\langle Y,\tilde Y\rangle = 1\\
&w^v_x n^u_x- w^u_x n^v_x  = 1\,,
\fe
where again we can satisfy the $SL(2,\mathbb{Z})$ condition since $\text{gcd}(w^u_x, w^v_x)=1$.
The two dual cycles $Y$ and $\tilde Y$ lead to a complete basis of $\Gamma$.
And as for $\tilde X$, we can redefine $\tilde Y$ by adding to it an arbitrary integer multiple of $Y$.

Other interesting intersections are
\ie\label{otherint}
&\langle \tilde X, \tilde Y\rangle = - n_y^u n_x^v+ n_y^v n_x^u\\
&\langle \tilde X, Y\rangle = - n_y^u w_x^v +n_y^v w_x^u \\
&\langle \tilde Y,  X\rangle = n_x^u w_y^v-n_x^v w_y^u ~.
\fe

The $\{X,\tilde X\}$ basis is related to the $\{Y,\tilde Y\}$ basis as
\ie\label{XtildeXYtildeY}
&\left(\begin{array}{c} X\\ \tilde X\end{array}\right)=\left(\begin{array}{cc} \langle X,\tilde Y\rangle & \,\, -\langle X,Y\rangle \\ \langle \tilde X,\tilde Y\rangle&\,\, -\langle \tilde X,Y\rangle\end{array}\right)\left(\begin{array}{c} Y \\ \tilde Y \end{array}\right)\,.
\fe
Since this is an $SL(2,\mathbb{Z})$ transformation, we have the identity
\ie\label{identity}
m\langle \tilde X,\tilde Y\rangle = 1-\langle \tilde X,Y\rangle\langle \tilde Y,X\rangle\,.
\fe

We limit ourselves to flat space with the obvious metric $ds^2=dx^2+dy^2$.  Then, the lengths of the $U$ and $V$ cycles and the angle between them are
\ie
&\ell_u=\sqrt{(\ell_x^u)^2 +(\ell_y^u)^2}\\
&\ell_v=\sqrt{(\ell_x^v)^2 +(\ell_y^v)^2}\\
&\cos \varphi = { \ell_y^u\ell_y^v+\ell_x^u\ell_x^v \over \ell_u\ell_v} ~.
\fe
The lengths of the   closed $X$ cycle and $Y$ cycle are
      \ie
      \label{fixedxl}
  &        \ell_x= w_y^u\ell_x^u-w_y^v\ell_x^v \,,\\
&     \ell_y= w_x^v\ell_y^v -w_x^u \ell_y^u \,.
     \fe

We also introduce the effective lengths
\ie\label{leffxyd}
&\ell^{\rm eff}_x\equiv{1\over m}\ell_x={\ell^u_x\over w^v_x}={\ell^v_x\over w^u_x} \,, \\
&\ell^{\rm eff}_y\equiv{1\over m}\ell_y ={\ell^u_y\over w^v_y}={\ell^v_y\over w^u_y}\,.
\fe
Note that the area of our torus can be expressed as
\begin{equation}
m \lxeff \lyeff = \ell_u \ell_v \sin \varphi = \ell_x^u\ell_y^v - \ell_y^u\ell_x^v \,.
\end{equation}

As we said above, it is convenient to replace the basis of cycles $\{U,V\}$ by $\{X,\tilde X\}$, i.e., to align the $U$ cycle with the $X$ cycle (see Figure \ref{XXtcycles}).  This corresponds to setting $\ell_y^u=0$ in \eqref{rotationxy} and leads to simplifications in some of the expressions above.  The modular parameter of our torus is then
\ie
&\tau=\tau_1+i\tau_2 = {k \over m} + i {\lyeff\over m\lxeff}\,,\\
&k \equiv \langle \tilde X, Y \rangle\,.
\fe
This makes it clear that our condition of finite wrapping amounts to $\tau_1={k \over m}$ being rational.  Here we also see that the independent data is $\lxeff$, $\lyeff$ and the two coprime integers $m$ and $k$.  In addition, the freedom mentioned above in shifting $k$ by a multiple of $m$ is recognized as being generated by the familiar $T$ transformation on $\tau$.

As we go around the $X$ and $\tilde X$ cycle, the coordinates $(x, y)$ are shifted as
\ie\label{xtxperiod}
&X:~ (x,y) \to ( x+m \lxeff  , y)\,,\\
&\tilde X:~(x, y)\to (x + k \lxeff, y + \lyeff )\,.
\fe

The geometric interpretation of the effective lengths is the following. Consider a periodic function on the torus that depends only on $y$.
The periodicity around the $\tilde X$ cycle \eqref{xtxperiod} means that
 \ie\label{xyperiodicity}
f(y) =  f(y + \ell^{\rm eff}_y)\,.
\fe
  Repeating this for a function $g(x)$ that depends only on $x$ we conclude that
\ie\label{kxky}
g(x) & = g(x+ \lxeff)\,,
\fe
i.e., their periodicities are smaller than $\ell_x$ and $\ell_y$.

The length of a closed contour along $y$ at fixed $x$ is given by $\ell_y =m \lyeff $, whereas the length of a closed contour along $x$ at fixed $y$ is given by $\ell_x = m \lxeff$. In other words,
\begin{equation}
\oint dy \equiv \int_{0}^{m \lyeff} dy\,,~~~\oint dx \equiv \int_{0}^{m \lxeff} dx\,.
\label{eq:closedcontours}
\end{equation}
Any well-defined function on our torus $f(x,y)$ must satisfy the periodicity constraints
\begin{equation}
\oint dy f(x,y) = \oint dy f(x + \lxeff,y)\,,~~~\oint dx f(x,y) = \oint dx f(x,y+\lyeff)\,.
\end{equation}

\begin{figure}
\centering
\includegraphics[width=120mm]{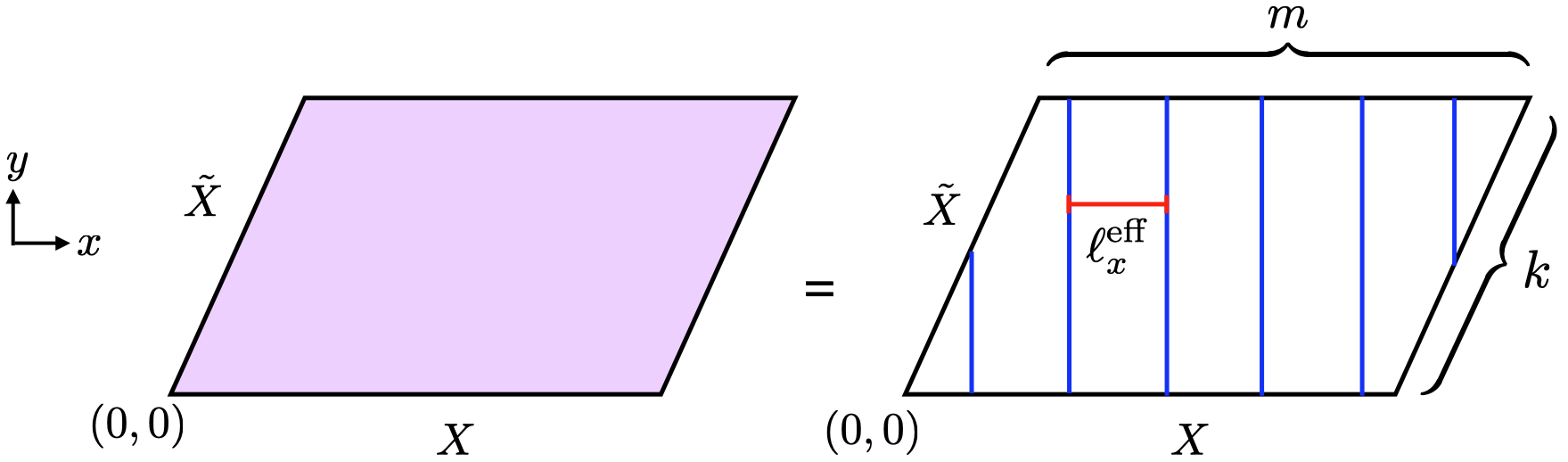}
\caption{The surface integral over the twisted torus in terms of the $x$ and $y$ integrals.  The contour of the integral $\oint dy$ runs along the blue lines, and the $\int dx$ integral runs over the red segment of length $\lxeff$.   }\label{twistedrec}
\end{figure}

To integrate a function $f(x, y)$ over the entire fundamental domain, we may first integrate over a closed contour at fixed $x$ and then integrate $x$ over a region of length $\lxeff$, or we may first integrate over a closed contour at fixed $y$ and then integrate $y$ over a region of length $\lyeff$. In particular, we have
\ie
\int_{T^2}dxdy f(x, y)&= \int_{0}^{\lxeff} dx \oint dy f(x, y) =\int_{0}^{\lyeff} dy \oint dx f(x, y)\, .
\fe Pictorially, the rewriting of the integral is shown in Figure \ref{twistedrec}.

\subsection{Lattice geometry}\label{ssec:Lattice}

We now consider a discretization of the twisted geometry by putting it on the lattice, whose sites are labeled by integers ($\hat x,\hat y$).
In particular, as in \eqref{rotationxy}, we consider identifications generated by
\begin{equation}\label{periodicityL}
(\hat x, \hat y) \sim (\hat x+L_x^u, \hat y + L_y^u) \sim (\hat x+L_x^v, \hat y + L_y^v) \,,
\end{equation}
with non-negative integers $L_i^r$.

As in the continuum discussion, we define integers $W_i^r$ describing the number of times a fixed $x$ or fixed $y$ curve runs around the cycles of our torus.  In terms of the parameters $L_i^r$ they are
\ie\label{integersd}
W_x^u \equiv  \frac{L_x^v}{\text{gcd}(L_x^u, L_x^v)}\,,~~W_x^v\equiv  \frac{L_x^u}{\text{gcd}(L_x^u, L_x^v)}\,,~~&W_y^u \equiv \frac{L_y^v}{\text{gcd}(L_y^u, L_y^v)} \,,~~W_y^v \equiv \frac{L_y^u}{\text{gcd}(L_y^u, L_y^v)}\,, \\
\gcd(W_x^v, W_x^u)= & \gcd(W_y^v, W_y^u) = 1 \,,
\fe
The lengths of the $X$ and $Y$ cycles are (compare with \eqref{fixedxl})
\ie
&L_x = W^u _y L ^u_x -W^v_y  L^v_x\,,\\
&L_y=W^v_x L^v_y-W^u_x L^u_y \,.
\fe

As in the continuum discussion \eqref{mdef}, we define
\ie
M \equiv  W_x^vW_y^u-W_x^uW_y^v\, .
\fe
The effective lengths of the $X$ and $Y$ cycles are
\ie\label{Leff}
& \Lxeff  \equiv {L_x\over M}  =\text{gcd}(L_x^u, L_x^v)  \,, \\
& \Lyeff  \equiv {L_y\over M}  = \text{gcd}(L_y^u, L_y^v)\,,
\fe
which are the lattice versions of continuum parameters $\lxeff$, $\lyeff$ of \eqref{leffxyd}.
They represent the periodicities of functions that depend only on $x$ or only on $y$.

As in the continuum description, it is convenient to use the basis of cycles $X$ and $\tilde X$ (see \eqref{XtildeXbasis})
\ie\label{XtildeXbasisl}
&\left(\begin{array}{c} X\\ \tilde X\end{array}\right)=\left(\begin{array}{cc} W_y^u & \, -W_y^v \\ - N_y^u&\, N_y^v\end{array}\right)\left(\begin{array}{c} U \\ V
\end{array}\right)\\
&W_y^uN_y^v-W_y^vN_y^u=1~.
\fe
These cycles correspond to
\ie\label{periodicityLX}
&X: (\hat x, \hat y)\to (\hat x+M\Lxeff, \hat y ) \\
&\tilde X: (\hat x, \hat y) \to (\hat x+ K \Lxeff, \hat y + \Lyeff)\,,\\
&K \equiv - N_y^u W_x^v +N_y^v W_x^u ~.
\fe

Next, we consider the continuum limit.
We introduce a lattice spacing $a$ and scale the integers $L_i^r$ such that the four limits
\ie
\ell_i^r = \lim_{a\to 0}a L_i^r
\fe
converge to their continuum counterparts.
Similarly,
\ie\label{simplelimit}
&\lim_{a\to 0} {W_x^v \over W_x^u} = {w_x^v\over w_x^u}\\
&\lim_{a\to 0} {W_y^v \over W_y^u} = {w_y^v\over w_y^u}\\
&\lim_{a \rightarrow 0} a \sqrt{ (L_x^u)^2 + (L_y^u)^2 }=\ell_u\\
&\lim_{a \rightarrow 0}   a \sqrt{ (L_x^v)^2 + (L_y^v)^2 }=\ell_v~.
\fe
However, the limits
\ie
\lim_{a\to 0} W_i^r\qquad, \qquad \lim_{a\to 0} M \qquad,\qquad \lim_{a\to 0} aL_i\qquad,\qquad \lim_{a\to 0} aL_i^\textrm{eff}
\fe
are not well-defined.  They do not necessarily converge to the continuum quantities $w_i^r$, $m$, $\ell_i$, $\ell_i^\textrm{eff}$.  They depend on the details of how we take $L_i^r$ to infinity.

As an extreme example of dependence on how we take the limit, consider two sequences of lattice geometries labeled by $L$, which we will take to infinity as $\ell/a$ with finite $\ell$.  The first is
\ie
L_x^u = L_y^v = L\,, \qquad L_x^v = L_y^u = 0\,.
\label{firstsequence}
\fe
and hence  $W_x^v=W_y^u=1$, $W_x^u=W_y^v=0$, $\Lxeff = \Lyeff = L$.  The second is
\ie
L_x^u = L_y^v = L\,, \qquad L_x^v = L_y^u = 1\,.
\label{secondsequence}
\fe
and hence $W_x^v=W_y^u=L$, $W_x^u=W_y^v=1$, $\Lxeff = \Lyeff = 1$.

As in \eqref{simplelimit}, both of them lead to the untwisted geometry $\ell_x^u=\ell_y^v=\ell$, $\ell_x^v=\ell_y^u=0$, ${w_x^u\over w_x^v} ={w_y^v\over w_y^u} =0$, $\ell_u=\ell_v=\ell$.  However, while the first one leads to finite $w_i^r =\lim_{a\to 0} W_i^r$, the second one leads to divergent $\lim_{a\to 0}W_i^r$, and misses the fact that the continuum values derived from  $\ell_x^u=\ell_y^v=\ell$, $\ell_x^v=\ell_y^u=0$ should be $w_x^v=w_y^u=1$, $w_x^u=w_y^v=0$.  Relatedly, it leads to $\lim_{a\to 0} L_i^\textrm{eff}=1$ and hence $\lim_{a\to 0} a L_i^\textrm{eff}=0$.

This example demonstrates also the discussion around \eqref{ellirc}.  The sequence \eqref{firstsequence} is the one that leads to a natural regularization of the continuum system.  Indeed, the continuum values of $w_i^r$ are the limits of the lattice values.

\section{Winding on the twisted two-torus}\label{sec:2+1winding}

In this section, we place circle-valued fields on our twisted torus.

As a warmup, let us start with a map from a one-dimensional circle of circumference $\ell$, which is parameterized by $x$ (i.e., $x \sim x+\ell$) to a target-space $f(x)$.
First, consider a smooth $f(x)$.
If $f$ is real-valued, then $f(x+\ell)=f(x)$.  If $f$ is circle valued, i.e., $f\sim f+2\pi$, then
\ie
e^{if(x+\ell)}=e^{if(x)}\,.
\fe
Lifting $f(x)$ to a real-valued function, we learn that $f(x+\ell)=f(x)+g$ with $g\in2\pi\bZ$.  We interpret $g$ as a transition function, which measures the winding number of the map ${g\over 2\pi}={1\over 2\pi} \oint dx \partial_x f(x)$.

Since we allow discontinuous $f$, this discussion should be modified.   We again lift $f$ to be real-valued. Then, we gauge $f(x)\sim f(x)+2\pi n(x)$ with $n(x)\in \bZ$, i.e., we allow an $x$ dependent, integer-valued gauge parameter $n(x)$.  Unlike the case of smooth $f$, where the lift at one point $x$ constrains the lift at nearby points, now there is no such constraint.  We can again consider a transition function $f(x+\ell)=f(x)+g(x)$ with $g(x)\in 2\pi \bZ$, but now we can choose another ``trivialization'' where $g(x)=0$, and therefore there is no winding number.  More explicitly, we can perform a non-periodic transformation $f(x) \to f(x) +2\pi n(x) $, $g (x)\to g(x)+2\pi(n(x+\ell)-n(x))$ to set $g(x)=0$.  (Note that locally this is a gauge transformation, but it changes the transition function because it is not periodic.)

Equivalently, as in \cite{paper1}, we can say that in this case $f$ and all its derivatives are not gauge invariant.  Only $e^{if}$ and its derivatives are gauge invariant.  Therefore, the winding charge ${1\over 2\pi}\oint dx \partial_x f$ is also not gauge invariant and it is not meaningful.

This discussion might appear as a fancy way of stating a well known fact.  When the circle parameterized by $x$ is a lattice and $f(x)$ is circle-valued, the configuration space does not break into sectors labeled by winding number -- there is no winding number on the lattice.  Nonetheless, our extended discussion here will prove quite useful below.

As preparation for later analysis, let us define some useful functions.  First, we will  use the periodic delta function
\ie\label{periodicd}
\delta^{\rm P}(x,x_0, \ell) =    \sum_{I\in \bZ} \delta(x-x_0 + I \ell )~.
 \fe
We will also find it convenient to define
\ie\label{ThetaPd}
\ThetaP(x,x_0, \ell_s)=\int_0^x dx'\delta^{\rm P}(x',x_0, \ell) ~.
\fe
Note that $\ThetaP(x,x_0, \ell)$ is not periodic.\footnote{Below we will sometimes  use $\ThetaP(x,0,\ell_s)$, which is subject to ambiguity given our convention that $\ThetaP(0,x_0,\ell_s)=0$ for any $x_0$.   To be more precise,  we will define $\ThetaP(x,0,\ell_s)$ as    $ \ThetaP (x,\epsilon,\ell_s)$ with $\epsilon$ positive and infinitesimally small.
}

\subsection{Transition functions and winding charges}\label{sec:transition}

We want to place a circle-valued field $\phi$, subject to the rules of \cite{paper1}, on the twisted torus.
In order to simplify the notation we will use the $SL(2,\mathbb{Z})$ freedom in redefining $U$ and $V$ and choose $U=X$ and $V=\tilde X$ from this point on.  Note that this choice also breaks the symmetry exchanging $X$ and $Y$ together with the other data characterizing these cycles.

As in \cite{paper1}, we will be interested in discontinuous functions $\phi$ with certain discontinuities.  Specifically, we allow $\phi$ to be discontinuous and therefore $\partial_x\phi$ and $\partial_y\phi$ can have delta-functions.  However, we restrict the discontinuities of $\phi$, such that $\partial_x\partial_y\phi$ can include a delta function in $x$ or in $y$, but we exclude situations where $\partial_x\partial_y\phi$ has terms like $\delta(x-x_0)\delta(y-y_0)$.  A special case of it is a discontinuous $\phi$ with finite  $\partial_x\partial_y\phi$.

We view the field $\phi$ as real-valued and to make it circle-valued by imposing the gauge identification
\ie\label{ngauge}
\phi(x,y) \sim \phi(x,y) + 2\pi \Big(n^x(x) +n^y(y)\Big) \qquad,\qquad n^x(x), n^y(y)\in  \mathbb{Z}~.
\fe
For an ordinary periodic scalar, the identification involves a position-independent integer.  Here  we allow discontinuous identifications of the form \eqref{ngauge}.  We do not include in the gauge identification an arbitrary integer valued function of both $x$ and $y$, as this takes us out of the space of functions we defined above.

We start with a real-valued field $\phi$ on $\mathbb{R}^2$.  We need to impose the gauge identification \eqref{ngauge} and place it on our torus $\mathbb{R}^2/\Gamma$.  Every vector in $\Gamma$ leads to a closed cycle $\cal C$ on our torus.  The identification across $\cal C$ should involve a transition function of the form \eqref{ngauge}
\ie\label{transitioncon}
&\phi(x+{\cal C}^x,y+{\cal C}^y) =\phi(x,y)+g_{\cal C}(x,y)\\
&g_{\cal C}(x,y)=2\pi \Big(n_{\cal C}^x(x)+ n_{\cal C}^y(y)\Big) \qquad,\qquad n_{\cal C}^x(x), n_{\cal C}^y(y)\in  \mathbb{Z}~.
\fe
Here $({\cal C}^x,{\cal C}^y) $ is the vector on the covering space corresponding to ${\cal C}$.  For example, for our basis of cycles $X$ and $\tilde X$,
\ie
&(X^x,X^y)=(m\lxeff,0)\\
&(\tilde X^x,\tilde X^y)=(k\lxeff,\lyeff)~
\fe
We will also discuss the $Y$ cycle, for which
\ie
(Y^x,Y^y)=(0,m\lyeff)~.
\fe

Our goal is to identify the distinct bundles.  This involves two steps.  First, we trivialize the bundle by choosing the transition functions.  Here we must impose the constraints from the cocycle conditions.  Second, we identify bundles labeled by different transition functions that are related by redefinitions.  Locally, these are gauge transformations, but globally they are not.

The composition of cycles ${\cal C}={\cal A}+{\cal B}$ leads to the cocycle condition
\ie\label{compositionc}
g_{\cal C}(x,y) = g_{\cal B}(x,y)+g_{\cal A}(x+{\cal B}^x,y+{\cal B}^y)= g_{\cal A}(x,y)+g_{\cal B}(x+{\cal A}^x,y+{\cal A}^y)~.
\fe

Using such a composition, it is enough to consider the transition functions for two generators of $\Gamma$, say $X$ and $\tilde X$, and express the other transition functions as linear combinations of these.  For example, the transition function of the $Y$ cycle is
\ie\label{gY}
&Y = - k X + m \tilde X\\
&g_Y (x,y)
=  - \sum_{I=1}^{k}   g_X (  x -  I m\lxeff,y  ) +\sum_{J=1}^{m}    g_{\tilde X} \left(  x -  J k \lxeff  ,y+(m-J)\lyeff\right)~.
\fe

The fact that the transition functions are separate functions of $x$ and $y$ \eqref{transitioncon} and the cocycle condition \eqref{compositionc} impose important constraints.  For example, the cocycle condition of the $X$ and $\tilde X$ cycles leads to
\ie\label{XtXcocycle}
n_X^y(y+\lyeff) - n_X^y(y) =
\left[n_{\tilde X}^x( x +m \lxeff)-n_{\tilde X}^x( x)  \right] - [ n_X^x( x  +k\lxeff)-n_X^x( x)]
\fe
and the cocycle condition of the $X$ and $Y$ cycles leads to
\ie\label{XYcocycle}
n_X^y(y+m\lyeff) - n_X^y(y) = n_{Y}^x( x +m \lxeff)-n_{Y}^x( x) ~.
\fe
From these two conditions and the equation obtained from \eqref{XtXcocycle} by exchanging $X$ and $Y$, we find the periodicity
\ie\label{XYperiodicity}
& g_X(x,y+\lyeff)=g_X(x,y)+  2\pi n_{xy}\\
&g_Y(x+\lxeff,y)= g_Y(x,y)+2\pi n_{xy}\,
\fe
with the same constant $n_{xy}$.

Next, we should identify bundles with different transition functions that are related by certain transformations.  Locally, these are gauge transformations, but they are not single valued.  Specifically, we identify
\ie\label{nonperiodic}
g_{\cal C}(x,y) \sim g_{\cal C}(x,y)+2\pi\Big(n^x(x+{\cal C}^x) - n^x(x)+n^y(y+{\cal C}^y)-n^y(y)\Big)~.
\fe
As a check, the cocycle condition \eqref{compositionc} is invariant under this identification.

Let us identify the  invariant information in the transition functions.
The action of the  transformation \eqref{nonperiodic} on the transition functions of the $X$ and $Y$ cycles implies that the $U(1)$ winding charges \eqref{windingci}
\ie\label{U1charges}
&Q^x(x)={1\over 2\pi}\oint dy \partial_x\partial_y\phi = {1\over 2\pi} \partial_x g_Y(x,y)=\partial_x n_Y^x(x)\,,\\
&Q^y(y)={1\over 2\pi}\oint dx \partial_x\partial_y\phi ={1\over 2\pi} \partial_y g_X(x,y)=\partial_y n_X^y(y) \,,
\fe
are  invariant.
Note that since $n_Y^x(x)$ and $n_X^y(y)$ are integers, the charges  are linear combinations of delta functions with integer coefficients. Furthermore, we have the periodicity \eqref{XYperiodicity}.
The integer $n_{xy}$ in \eqref{XYperiodicity} can now be interpreted as a constant used in \cite{paper1}:
\ie\label{nxy}
\int_0^{\lxeff} dx Q^x(x) = \int_0^{\lyeff} dy Q^y(y)=  n_{xy}~.
\fe

So far we have identified a continuum of $U(1)$ charges $Q^x(x)$ and $Q^y(y)$ labeled by $0\le x<\lxeff , 0\le y<\lyeff$, subject to the constraint \eqref{nxy}.
If we regularize the theory on a lattice, it leads to $\Lxeff+\Lyeff-1$ integer charges.

In addition to these integers, we have a $\bZ_m$-valued phase:
\ie\label{Zmcharge}
{\cal U}(x,y) &= \exp\left[
- {i\over m } \sum_{I=0}^{m-1}  g_{\tilde X}( x+I \lxeff,y)   + {i\over m} \sum_{J=0}^{k-1} g_X(x+J\lxeff,y)
\right]\,.
\fe
We will motivate this operator and discuss it further in Appendices \ref{app:counting} and \ref{app:Uphi0}.
It is straightforward to check that it is  invariant under \eqref{nonperiodic}.
Naively this defines many $\bZ_m$ charges depending on the choice of $x,y$.
However, using \eqref{gY} and the cocycle conditions, we have
\ie
{{\cal U}(x,y)\over {\cal U}(0,0)} =
\exp\left[ -{2\pi i \over m}  \int_0^x dx' Q^x(x') \right]
\exp\left[ {2\pi k i \over m}  \int_0^y dy' Q^y(y') \right] \,,
\fe
which is a function of $Q^x(x)$ and $Q^y(y)$.
Therefore,  ${\cal U}(x,y)$ leads to a single  $\bZ_m$ invariant beyond the $U(1)$ charges $Q^x(x),Q^y(y)$.

This $\bZ_m$ charge can also be written directly in terms of $\phi$:
\ie\label{Zmchargephie}
{\cal U}(x,y) &= \exp\left[
- {i\over m} \sum_{I=0}^{m-1} \left(
\, \phi\left(x+(k+I)\lxeff , y+\lyeff\right) -\phi\left (x+(k+I)\lxeff,y\right) \,
\right)
\right] \,.
\fe

We refer the readers for a more detailed discussion on related points to the appendices.
In Appendix \ref{app:counting}, we will verify the  number of winding charges by classifying all  the invariants of the transition functions.
In  Appendix \ref{app:Uphi0}, we will discuss  additional operators in the $\phi$-theory and motivate  the $\bZ_m$ charge \eqref{Zmcharge}.

We can summarize this discussion as follows.  Windings around the $X$ and $Y$ cycles are measured by the charges $Q^y(y)$ and $Q^x(x)$.  They are essentially the same as the windings in the untwisted torus, except that they have periodicities $\lxeff$ and $\lyeff$ respectively.  The new charge $\cal U$ is present because our torus has additional cycles.  As stated after \eqref{hatY}, if we mod out by the $X$ and $Y$ cycles, the $\tilde X$ cycle behaves like a torsion cycle.  This leads to the fact that the new charge is a $\bZ_m$ charge.

\subsection{Winding configurations of $\phi$}\label{sec:2+1windingphi}

In this section, we present  winding configurations of $\phi$ that realize the winding charges in Section \ref{sec:transition}.

\subsubsection{Special configurations}\label{sec:special}

We start with the winding configurations satisfying
\ie\label{phizde}
\partial_x\partial_y\phi=0\,.
\fe
These special configurations will be useful for other discussions below.

As a warmup, let us start with a real-valued function on the torus.  In \eqref{xyperiodicity} and \eqref{kxky}, we studied a real-valued function that depends only on $x$ or only on $y$ and found that it has periodicity $\lxeff$ and $\lyeff$ respectively.  A trivial extension of this analysis applies to the case of a real-valued function $f$ satisfying $\partial_x\partial_y f=0$.  Because of the differential equation, we have $f(x,y)=f^x(x)+f^y(y)$, and the boundary conditions set
\ie\label{fxyp}
&f^x(x+\lxeff)=f^x(x)\,, \\
&f^y(y+\lyeff)=f^y(y)\,.
\fe

We will see that the conclusion is different for a circle-valued field $\phi\sim\phi+2\pi$.
Locally, we can solve \eqref{phizde} as
\ie\label{phizd}
\phi(x,y)=\phi^x(x)+\phi^y(y)\,.
\fe
The boundary conditions tell us that
\ie
&e^{i\phi^x(x+m\lxeff)} =e^{i\phi^x(x)}\,,\\
&e^{i\phi^x(x+k\lxeff)}e^{i\phi^y(y+\lyeff)} =e^{i\phi^x(x)}e^{i\phi^y(y)}\,.
\fe
This means that
\ie\label{phizxy}
&e^{i\phi^x(x+\lxeff)}=\eta e^{i\phi^x(x)}\\
&e^{i\phi^y(y+\lyeff)}=\eta^{-k} e^{i\phi^y(y)}\\
&\eta^m=1
\fe
with a position-independent $\eta$.

We see that while a real-valued function $f$ has the simple periodicity \eqref{fxyp}, a circle-valued function $\phi$ has a new $\bZ_m$ phase $\eta$ in that periodicity \eqref{phizxy}.

The most general such $\phi$ can be expressed as
\ie\label{generalphiz}
&\phi(x,y)={2\pi r\over m}\left({x\over \lxeff}-k {y\over \lyeff}\right) +f^x(x)+f^y(y)\,,\\
&e^{if^x(x+\lxeff)}=e^{if^x(x)}\,,~~~~e^{if^y(y+\lyeff)}=e^{if^y(y)}\,,\\
&r=1,2,...,m\,.
\fe
It carries a nontrivial $\bZ_m$ charge \eqref{Zmcharge}, ${\cal U}(x,y) = e^{2\pi i kr/m}$, but zero $U(1)$ charges, $Q^i(x^i)=0$.

As in \cite{paper1}, we  are also  interested in discontinuous functions with certain discontinuities.
In particular,  $f^x(x)$ and $f^y(y)$ in \eqref{generalphiz} can be discontinuous.
Also, the field $\phi$ is subject to a discontinuous gauge transformation \eqref{ngauge}.

This has two important consequences. First, we can replace the first term in \eqref{generalphiz} by another function with the same transition functions, e.g.
\ie\label{generalphiz2}
\phi(x,y)={2\pi r\over m}\left(\ThetaP(x,0,\lxeff)-k \ThetaP(y,0, \lyeff)\right) +f^x(x)+f^y(y)
\fe
with different $f^x$ and $f^y$.  Second, as in the discussion of the one-dimensional case above, the fact that $ e^{if^x(x+\lxeff)}=e^{if^x(x)}$ and $e^{if^y(y+\lyeff)}=e^{if^y(y)} $ means that we can choose a lift where $ f^x(x+\lxeff)=f^x(x)$ and $f^y(y+\lyeff)=f^y(y) $ as real functions.

\subsubsection{More general configurations}

Next, we consider configurations  that carry nontrivial $U(1)$ charges with  $\partial_x\partial_y\phi\neq 0$.
The  minimal winding configuration with nontrivial $Q^x(x)$ and $Q^y(y)$ should satisfy
\ie\label{twodch}
Q^x(x)=& {1\over 2\pi} \oint dy \partial_x\partial_y \phi = \delta^{\rm P}(x,x_0, \lxeff)\,,\\
Q^y(y)=& {1\over 2\pi} \oint dx \partial_x\partial_y \phi = \delta^{\rm P}(y,y_0, \lyeff)\,,
\fe
for some $x_0$ and $y_0$.
The periodic delta function $\deltaP(x,x_0,\lxeff)$ was defined in \eqref{periodicd}.
These configurations can also carry the $\bZ_m$ charge \eqref{Zmcharge}.

The charges \eqref{twodch} lead us to look for a configuration satisfying
\ie\label{ppphi}
\partial_x\partial_y \phi(x,y) =  {2 \pi\over m} \Bigg[ & \frac{1}{\lxeff}  \delta^{\rm P}(y,y_0, \lyeff)
 + \frac{1}{\lyeff} \delta^{\rm P}(x,x_0, \lxeff)
-  \frac{1}{\lxeff \lyeff}  \Bigg]
\fe
for some $x_0$ and $y_0$.
Such a  minimal winding configuration is given by
\ie\label{minimalphi}
& \phi = 2\pi
\left( {x\over \ell_x} -\langle\tilde X,Y\rangle {y\over \ell_y}\right) \, \ThetaP(y,y_0 ,\lyeff)
 + 2\pi \left( {y\over \ell_y} - \langle\tilde Y,X\rangle {x\over \ell_x} \right) \, \ThetaP(x,x_0 ,\lxeff)\\
&
- {2\pi m}
 \left[ - \frac12\langle \tilde X,Y\rangle  \left({y \over \ell_y}\right)^2  - \frac12\langle \tilde Y,X\rangle \left({ x\over \ell_x}\right)^2 +{xy\over \ell_x\ell_y } \right]
 +2\pi \left(c_x {x\over \ell_x}  +c_y { y\over \ell_y }\right)\\
&c_x = {m\over 2}\langle \tilde Y,X\rangle +r \,,~~~~~c_y = -{m\over 2}\langle\tilde X,Y\rangle  \left( \langle \tilde X,\tilde Y\rangle +\langle\tilde Y,X\rangle\right)  -  \langle\tilde X,Y\rangle r\,.
\fe
Here we have chosen  $c_x,c_y$ such that the transition functions   take  a simple form (see below).
We also have the freedom of adding a standard winding configuration \eqref{generalphiz} to $\phi$ and shifting the value of $r$.

Let us check that the transition functions are indeed $2\pi \bZ$-valued.
Using \eqref{compositionc}, it suffices to check this for the transition functions for the $X$ and $\tilde X$ cycles.
The transition function around the $X$ cycle is
\ie
&g_X  (x,y)  = \phi \left(x+m\lxeff \, ,\, y \right) -\phi(x,y)=   2\pi \ThetaP(y,y_0,\lyeff)- 2\pi \langle \tilde Y,X\rangle\ThetaP(x,x_0,\lxeff)  +2\pi r\,.
\fe
Similarly, the transition function around the $\tilde X$ cycle is
\ie
&g_{\tilde X} (x,y) =\phi\left(x+\langle\tilde X,Y\rangle \lxeff\, ,\, y+ \lyeff \right)  - \phi(x,y)= 2\pi \langle \tilde X,\tilde Y\rangle  \ThetaP(x,x_0,\lxeff) \,.
\fe
Using \eqref{identity}, one finds that the cocycle condition is satisfied.

In addition to the winding charges $Q^x(x),Q^y(y)$, the minimal winding configuration \eqref{minimalphi} also carries the $\bZ_m$ charge \eqref{Zmcharge}
\ie
{\cal U}(x,y)
&=
\exp\left[ -{2\pi i \over m}  \int_0^x dx Q^x(x) \right]
\exp\left[ {2\pi k i \over m}  \int_0^y dy Q^y(y) \right]\\
&\quad\times \exp\left[
{2\pi i \over m} \left(
 k r +r_0
\right)
\right]
\fe
where $r_0=- {(m-1)m\over2}\langle\tilde X,\tilde Y\rangle  -{(k-1)k\over 2}  \langle\tilde Y,X\rangle  $.
The first line, which depends on $x,y$, is expressed in terms of the the winding charges $Q^x(x),Q^y(y)$.

\subsection{Comments about the $2+1$-dimensional $\phi$-theory}\label{sec:commentphi}

The $2+1$-dimensional $\phi$-theory \eqref{phiti}, which had been introduced in \cite{PhysRevB.66.054526}, was studied in \cite{paper1}  on an untwisted torus. Its main features are
\begin{itemize}
\item The theory has ``momentum'' and ``winding'' subsystem symmetries, \eqref{phicop} and \eqref{windingJ},  each of which leads (on the lattice) to $L_x+L_y-1$ conserved $U(1)$ charges.
\item In the quantum theory, all the states charged under these symmetries acquire large energy, of the order of the UV cutoff.
\item The theory is self-dual.  The duality exchanges the original field $\phi$ with another field $\phi^{xy}$.  It also exchanges the momentum and winding symmetries.
\end{itemize}

Let us see how this picture changes on the twisted torus.  First, it is clear that the conserved momentum and winding currents remain conserved.  By analogy to the untwisted case, we now have $\Lxeff+\Lyeff-1$ conserved $U(1)$ charges.  It is also clear that all the states charged under these symmetries acquire large energy in the quantum theory.

The main novelty in the problem on the twisted torus is associated with the configurations
\eqref{generalphiz} with $r\ne 0$  mod $m$.
These configurations have two consequences.  First, they carry a discrete $\mathbb{Z}_m$ winding charge under
\eqref{Zmcharge}.
Since this operator can be defined in terms of the transition functions, it is a conserved operator in the $\phi$-theory.
See Appendix \ref{app:Uphi0} for more discussion on the winding operator.

Second, a shift of $\phi$ by \eqref{generalphiz} is a momentum symmetry.  Clearly, it is not included in the $L_x^{\rm eff}+L_y^{\rm eff}-1$  $U(1)$ momentum symmetries.  Instead, this shift amounts to a $\mathbb{Z}_m$ momentum symmetry. This symmetry operator cannot be written simply in terms of the $\phi$ field -- it is a twist   operator of $\phi$.
 Alternatively, it is represented, as in \eqref{Zmchargephie}, in terms of the dual field $\phi^{xy}$ as
\ie\label{Zmchargephiexy}
{\cal V}(x,y) = \exp\left[
-{i\over m} \sum_{I=0}^{m-1} \left(
\, \phi^{xy}\left(x+(k+I)\lxeff , y+\lyeff\right) -\phi^{xy}\left (x+(k+I)\lxeff,y\right) \,
\right)
\right] \,.
\fe
This operator has all the properties of ${\cal U}(x,y)$ that we mentioned above.  In particular, up to adding momentum charges, it is independent of $x$ and $y$.

The two $\mathbb{Z}_m$ global symmetries generated by \eqref{Zmchargephie} and \eqref{Zmchargephiexy} commute with all the $U(1)$ momentum and winding symmetries.
 But they do not commute with each other.  They generate a clock and shift algebra
\ie
&{\cal U}^m={\cal V}^m=1\,,\\
&{\cal U}{\cal V}=e^{-2\pi i{k \over m}}{\cal V}{\cal U}~.
\fe
(Since $\text{gcd}(k,m)=1$, we can redefine the generators of this algebra to make the phase above $e^{2\pi i /m}$ as in \eqref{ZmZm}.)
This algebra has an $m$-dimensional representation.  Therefore, the Hilbert space of our problem includes a factor of this $m$-dimensional representation.  In particular, the system must have $m$ degenerate ground states.

We conclude that unlike the theory on the untwisted torus, here we have two  $\mathbb{Z}_m$ symmetries, and these two symmetries do not commute.  Clearly, these two symmetries are exchanged under the self-duality of the system.  Also, unlike the $U(1)$ momentum and winding symmetries, all the states in the theory transform under these symmetries in their $m$-dimensional representation.
As we mentioned in the introduction, these effects are reminiscent of the phenomena discovered in \cite{Freed:2006ya, Freed:2006yc}.

\section{$2+1$-dimensional $\mathbb{Z}_N$ tensor gauge theory}\label{sec:2+1d}

As a warm-up for the $3+1$-dimensional X-cube model, we start by placing the $2+1$-dimensional  $\bZ_N$ tensor gauge theory in \cite{paper1} on a two-dimensional spatial torus with twisted boundary condition \eqref{rotationxy}.
We will analyze this model using the two dual presentations in \cite{paper1}.

\subsection{Special $\bZ_N$ configurations}\label{sec:specialZN}

In the rest of this paper, we will frequently consider a $\bZ_N$-valued $\phi$ field on a twisted torus.
More explicitly, such a $\bZ_N$ field $\phi$ obeys the same rule as in Section \ref{sec:transition} plus one $\bZ_N$ condition:
\ie
 {N\phi\over 2\pi} \in \bZ\,.
 \fe
Similar to Section \ref{sec:special}, of particular interest is the special case when
\ie
\partial_x\partial_y\phi=0,
\fe
is obeyed.

The most general expression of such a $\bZ_N$-valued $\phi$ can be found through an analysis similar to that of Section \ref{sec:special}.
 The phase $\eta$ there now obeys not only $\eta^m=1$, but also $\eta^N=1$.  Hence it is a $\bZ_{\text{gcd}(N,m)}$ phase. In conclusion, the most general such $\phi$ takes the form:
 \ie\label{generalZNphi}
\phi=  {2\pi r\over \text{gcd}(N,m)} \left( \ThetaP(x,0,\lxeff)   - k \ThetaP(y, 0,\lyeff)  \right)
+ {2\pi \over N} \left( W^x(x) +W^y(y) \right)\,,
\fe
where $r=1,\cdots,\text{gcd}(N,m)$ and $W^i(x^i)\in \bZ$.
Using the freedom in  \eqref{ngauge} to redefine $\phi$, we can choose a lift of $W^i(x^i)$ such that $W^i(x^i+\ell_i^{\rm eff})= W^i(x^i)$.

\subsection{$\phi-A$ theory}\label{sec:2+1phiA}

The first presentation of the $\bZ_N$ model is based on the Lagrangian
\ie\label{ZNLag}
{\cal L} =
 {1 \over 2\pi} \widetilde E^{xy} ( \partial_x \partial_y \phi - NA_{xy})
+ {1\over 2\pi }\widetilde B(\partial_0\phi - NA_0) \,,
\fe
where  $(A_0,A_{xy})$ are  $U(1)$ tensor gauge fields and $\phi$ is a  $2\pi$-periodic real scalar field that Higgses the $U(1)$ gauge symmetry to $\bZ_N$.
The  gauge transformations of these fields are
\ie
&\phi \sim \phi + N  \alpha\,,\\
&A_0\sim A_0  + \partial_0\alpha\,,\\
&A_{xy} \sim A_{xy} +\partial_x\partial_y \alpha\,.
\fe
The fields $\widetilde E^{xy}$ and $\widetilde B$ are Lagrange multipliers. Their coefficients are are not important at this stage, but we set them such that $\widetilde  E^{xy}$ and $\widetilde B$ are standardly normalized field strengths in the dual picture.
The equations of motion are
\ie\label{eomZN1}
&\partial_x \partial_y \phi - NA_{xy}=0\,,\\
&\partial_0\phi  - NA_0 =0\,,\\
&\widetilde E^{xy}= \widetilde B=0\,.
\fe

Using the equations of motion \eqref{eomZN1}, we can solve all the other fields in terms of $\phi$. Then, we mod out by gauge transformations $\phi\sim \phi+N\alpha$.  The remaining configurations are linear combinations of the winding mode \eqref{minimalphi} with different $x_0,y_0$.  The coefficients in the linear combinations are   in the set $\{0,1,...,N-1\}$.
In addition, the winding configuration is labeled by an integer $r$, which in the $\bZ_N$ theory is defined  modulo gcd$(N,m)$.

We will regularize the ground state degeneracy by putting the theory on a lattice.
As discussed in Section \ref{sec:GEOMETRY}, the discretization of a continuum geometry is not unique, and we will see that the ground state degeneracy depends not only the continuum geometric data $\ell_i^r$, but also the details of the lattice regularization.

Let us consider a lattice geometry of the form discussed in Section \ref{ssec:Lattice}.
From this point on, the analysis of the ground state degeneracy proceeds in an analogous way as in \cite{paper1}, with the replacement $L_x\to L_x^{\rm eff},L_y\to L_y^{\rm eff}$.
Recall that  $L_x^{\textrm{eff}} =  \text{gcd}(L_x^u, L_x^v)\,,~L_y^{\textrm{eff}} =  \text{gcd}(L_y^u, L_y^v)$.
A general winding configuration on this lattice is labeled by a choice of integers $W^x_\alpha \in \{ 0, 1,...,N-1\}$ for each $\hat x_\alpha=1,\cdots ,L_x^{\rm eff}$ on the lattice and $W^y_\beta \in \{ 0, 1,...,N-1\}$ for each $\hat y_\beta=1,\cdots ,L_y^{\rm eff}$ on the lattice.
We also have the constraint
\ie
\sum_{\alpha}  W^x_\alpha= \sum_{\beta}W^y_\beta \,.
\fe
In addition to these integers, the winding configuration is further labeled by a $\bZ_{\text{gcd}(N,M)}$ -valued integer $r$. (Recall that $M$ is the lattice version of $m$.)
Combining these together,  the ground state degeneracy of our model is given by
\ie\label{2+1GSD}
N^{L_x^{\textrm{eff}}+L_y^{\textrm{eff}}-1} \,\text{gcd}(N,M)\, .
\fe

This formula for the ground state degeneracy has several peculiar features. Like that of the untwisted model in \cite{paper1}, the logarithm of the ground state degeneracy grows with the size of the system in a sub-extensive manner. In contrast with that of the untwisted model, however, this ground state degeneracy does not vary monotonically under small changes in the parameters $L_i^r$. Relatedly, it does not have a well-defined continuum limit.
To see this, let us compare the ground state degeneracy of two sequences of lattice models with the same continuum limit.  In particular, consider the sequence in \eqref{firstsequence},
\ie
L_x^u = L_y^v = L\,, \qquad L_x^v = L_y^u = 0\,,
\fe
and the sequence in \eqref{secondsequence},
\ie
L_x^u = L_y^v = L\,, \qquad L_x^v = L_y^u = 1\,.
\fe
These sequences both approach the same continuum quantities $\ell^r_i$.
But the first of these sequences has $\Lxeff = \Lyeff = L$, and $M=1$, and hence a ground state degeneracy of $N^{2 L -1}$, whereas the second  sequence has $\Lxeff = \Lyeff = 1$, and $M=L^2-1$, and hence a ground state degeneracy of $N \text{gcd}(N,L^2-1)$.  The ground state degeneracy of these models is therefore completely different: the first diverges in the continuum limit $L \rightarrow \infty$, whereas the second is ill-defined.  As we said above, the first sequence is the natural choice for this continuum theory.

\subsubsection{Using transition functions}

In the previous analysis, as in the discussion of this theory on the untwisted torus in \cite{paper1}, we assumed that we can always set the transition functions of the gauge theory on the spatial two-torus to be trivial.  Here we will show that the same conclusion is obtained by allowing arbitrary transition functions.

We consider nontrivial circle-valued transition functions $\gamma_{\cal C}$ that determine
\ie\label{AphitrP}
&A_{xy}(x+{\cal C}^x, y+{\cal C}^y)=A_{xy}(x,y) +\partial_x\partial_y \gamma_{\cal C}(x,y)\,\\
&\Phi(x+{\cal C}^x, y+{\cal C}^y)=e^{iJ\gamma_{\cal C}(x,y)} \Phi(x,y)\,
\fe
where $\Phi$ is a complex field with charge $J$.
(We will limit ourselves to static configurations.)
The composition  of cycles ${\cal C}={\cal A}+{\cal B}$ leads to the cocycle condition:
\ie\label{hatcocycle}
&\exp\left[ i   \gamma_{\cal B} (x,y)  + i \gamma_{\cal A} (x+{\cal B}^x ,y+{\cal B}^y  )\right]=\exp\left[i   \gamma_{\cal A}  (x,y)  + i \gamma_{\cal B} (x+{\cal A}^x ,y+{\cal A}^y  )\right]\,.
\fe

Next, we identify configurations with different transition functions that are related by certain transformations.
Specifically, for any circle-valued function $e^{i\beta}$, we identify
\ie\label{beta}
&\exp[i\gamma_{\cal C} (x,y)] \sim \exp\left[
i  \gamma_{\cal C} (x,y) +i\beta(x+{\cal C}^x ,y+{\cal C}^y) -i\beta(x,y)
\right]\,,\\
&A_{xy} (x,y) \sim A_{xy}(x,y) +\partial_x\partial_y \beta(x,y)\,\\
&\Phi(x, y)\sim e^{iJ\beta(x,y)} \Phi(x,y)\,.
\fe
If $e^{i\beta}$ is single-valued on our torus, then this is a gauge transformation, and it does not change the transition functions. Otherwise, it relates different trivializations of the same configuration.

Consider first the pure gauge $A$-theory.  Locally, we can choose $A_{xy}=0$ and then all the information about the gauge fields is in the transition functions.
The analysis of these transition functions is parallel to the discussion of the transition functions and winding configurations in the $\phi$-theory in Section \ref{sec:transition} and Appendix \ref{app:counting}.  There is only one difference: the integer valued functions $g_{\cal C}$ of the scalar theory are replaced in the gauge theory with real, circle-valued functions $\gamma_{\cal C}$.

First, we focus on the $X$ and $Y$ cycles.  We will return to the $\tilde X$ cycle shortly.  We find
\ie\label{Atheoryt}
&e^{i \gamma_X}=e^{i(f_X^x(x)+f_X^y(y))}\,, \\
&e^{i \gamma_Y}=e^{i(f_Y^x(x)+f_Y^y(y))}\,, \\
&e^{if_Y^x(x+\lxeff)}=e^{if_Y^x(x)+ i\varphi}\, , \\
&e^{if_X^y(y+\lyeff)}=e^{if_X^y(y)+i\varphi} \,.
\fe
As in in Section \ref{sec:transition} and Appendix \ref{app:counting}, $\partial_y f_X^y(y)$,  $\partial_x f_Y^x(x)$, and $\varphi$ are physical and gauge invariant.  (We will soon relate them to the holonomy around the $X$ and $Y$ cycles.)  In order to check whether there is additional invariant information, we follow the approach in Appendix \ref{app:counting} and set these quantities to zero and look for more data.  In particular, we look for additional information in
\ie\label{Atheoryt}
&e^{i \gamma_X}=e^{if_X^x(x)} \, , \\
&e^{i \gamma_Y}=e^{if_Y^y(y)} \, , \\
&e^{i \gamma_{\tilde X}}=e^{i(f_{\tilde X}^x(x)+f_{\tilde X}^y(y))} \, .\\
\fe
Imposing the cocycle conditions and using the freedom to change the trivialization, we find that all the functions here can be set to zero.

Unlike the analysis of the $\phi$-theory, there is no additional $\bZ_m$ charge.  Specifically, in following the steps in Appendix \ref{app:counting} with circle-valued functions, rather than integer-valued functions, the identification \eqref{scriptN} becomes
\ie\label{scriptZ}
&e^{if^x_{\tilde X}(x) }\sim e^{i (f^x_{\tilde X}(x) +  C _{\tilde X})}\,,\\
&e^{if^y_{\tilde X}(y)}\sim e^{i (f^y_{\tilde X}(y)  -  C_{\tilde X})}\,,\\
&e^{i{\cal Z}}\sim e^{i({\cal Z} -  m C _{\tilde X})}\,,
\fe
where $C _{\tilde X}$ and ${\cal Z}$ were denoted in Appendix \ref{app:counting}  by $N _{\tilde X}$ and $\cal N$, respectively.  Since now they are circle-valued, we can set $e^{i{\cal Z}}=1$ and there is no additional $\bZ_m$ charge.

We end up with the same data we have with trivial transition functions, but with nonzero
\ie\label{Axye}
A_{xy}= {1\over \ell_y }  f^x(x)  +{1\over \ell_x }  f^y(y)\,,
\fe
with $\partial_x f_Y^x(x)=f^x(x)+{1\over \ell_x} \oint_Y dy f^y(y)$, $\partial_y f_X^y(y)=f^y(y)+{1\over \ell_y} \oint_X dx f^x(x)$.  As a check, they both have the same holonomies
\ie\label{Aholonomies}
&W_X=\exp\left(i\int_{y_1}^{y_2} dy \oint_X dx  \left({1\over \ell_y }  f^x(x)  +{1\over \ell_x }f^y(y) \right)\right) =e^{i ( \gamma_X(x,y_2)-\gamma_X(x,y_1) ) } \,, \\
&W_Y=\exp\left(i\int_{x_1}^{x_2} dx \oint_Y dy  \left({1\over \ell_x }  f^y(y)  +{1\over \ell_y }f^x(x) \right)\right) =e^{i ( \gamma_Y(x_2,y)-\gamma_Y(x_1,y) ) }\,.
\fe

Let us repeat this analysis in the $\bZ_N$ theory using this perspective of the Higgs theory \eqref{ZNLag}.
The matter field $\phi$ transforms such that $\Phi=e^{i\phi}$ in \eqref{AphitrP} and \eqref{beta} has charge $J=N$.
We choose the unitary gauge $\phi=0$ and set $A_{xy}=0$.  In order for the gauge choice $\phi=0$ to be meaningful, the transition functions \eqref{Atheoryt} should be $N$'th roots of unity.
In contrast to the $U(1)$ gauge theory of $A$, we can no longer use the identification \eqref{scriptZ} to set $e^{i{\cal Z}}=1$.  In more detail, now $e^{i {\cal Z}}$ and $e^{i C_{\tilde X}}$ in \eqref{scriptZ} are $N$'th roots of unity.  Therefore, \eqref{scriptZ} identifies $e^{i{\cal Z}} \sim e^{i ( {\cal Z}-  {2\pi m\over N})}$
and we end up with $\gcd(N,m)$ distinct values.
Placing this result on the lattice, we reproduce \eqref{2+1GSD}.

Let us phrase it more explicitly.  The $\bZ_N$ theory has $\text{gcd}(N,m)$ configurations:
\ie\label{AZNe}
&e^{ i \phi}=1\,,~~~~A_0=A_{xy}=0\,,\\
&e^{i\gamma_X} = e^{ 2\pi i  r {m\over N\text{gcd}(N,m)}}\,,~~~~e^{i\gamma_{\tilde X}}=1\,,
\fe
with $r=0,1,\cdots, \text{gcd}(N,m)-1$.  Such configurations are present in the $U(1)$ $A$-theory, but they do not contribute to the holonomies \eqref{Aholonomies}.  In fact, they are identified with the trivial configuration with $e^{i\gamma_X}=1$ by a change of the trivialization \eqref{beta}, with e.g.,
\ie\label{ZNbeta}
e^{i\beta} = \exp\left[ -{2\pi i  r \over N\text{gcd}(N,m)  }  \left({x\over\lxeff} - k {y\over \lyeff} \right)  \right]
\fe
or
\ie\label{ZNbetas}
e^{i\beta} = \exp\left[ -{2\pi i  r \over N\text{gcd}(N,m)  }  \left(\ThetaP(x,0,\lxeff) - k\ThetaP(y,0,\lyeff) \right)  \right]
\fe
Therefore, the $U(1)$ theory does not have another label associated with these configurations.  This is to be contrasted with the situation in the $\bZ_N$ theory.  Here, as we explained above,we cannot perform identifications like \eqref{ZNbeta} or \eqref{ZNbetas} because they are not $\bZ_N$-valued.  (Equivalently, they do not preserve the choice $\phi=0$ in the Higgs theory \eqref{ZNLag}.)
Consequently, the configurations \eqref{AZNe} are nontrivial in the $\bZ_N$ theory and lead to the factor $\text{gcd}(N,M)$ in \eqref{2+1GSD}.

Let us offer a broader view on the analyses of the transition functions in the various theories.
The transition functions for the $\phi$-theory in Section \ref{sec:transition}, and those for the $U(1)$ and $\bZ_N$ tensor gauge theories in this section, are subject to similar cocycle conditions and identifications, but with coefficients valued in different groups,  $\bZ$, $U(1)$, and $ \bZ_N$ respectively.
In these three cases, there is an additional $\bZ_m$ label for the $\phi$-theory, no such label for the $U(1)$ $A$-theory, and an additional $\bZ_{\text{gcd}(N,m)}$ label for the $\bZ_N$-theory.
These additional labels can be thought of as torsion parts of an appropriate cohomology with $\bZ$, $U(1)$, and $ \bZ_N$ coefficients.

\subsection{$BF$-type $\bZ_N$ tensor gauge theory}\label{sec:2+1BF}

Next, we compute the ground state degeneracy using a dual presentation of the same $\bZ_N$ model \cite{paper1}:\footnote{Since $\phi^{xy}$ is circle-valued, the Lagrangian has to be defined more carefully.  Specifically, we can choose a trivialization, use this expression in each patch, and add correction terms similar to those in Appendix \ref{app:Uphi0}, in the overlap regions.  We will not do it here.}
\ie\label{ZNLag2}
{\cal L} =
{1 \over 2\pi} N \phi^{xy}(\partial_0 A_{xy} - \partial_x\partial_y A_0)= {1 \over 2\pi}N \phi^{xy}E_{xy}\,.
\fe

The phase space is given in the temporal gauge $A_0=0$ by
\ie
\left\{
\phi^{xy}(x,y) , A_{xy} (x,y) ~\Big| ~
\partial_x\partial_y \phi^{xy}=0\,,~~A_{xy}(x,y) \sim A_{xy} (x,y)+ \partial_x \partial_y \alpha(x,y)\right\}\,.
\fe
This is solved modulo gauge transformations by\footnote{Here we take the transition functions for $A_{xy}$ to be trivial.  Alternatively, as above, we can set $A_{xy}=0$ and have the nontrivial information in the transition functions.}
\ie\label{phiApart}
&A_{xy}  = {1\over \ell_y }  f^x(x)  +{1\over \ell_x }  f^y(y) \,,\\
&\phi^{xy} =  {2\pi r\over m } \left( \ThetaP(x,0,\lxeff) -  k \ThetaP(y,0,\lyeff)\right) +   \hat f_x (x) +\hat f_y (y) \,,
\fe
where   $r=1,\cdots , m$.  Here the functions $f^i(x^i)$ and $\hat f_i(x^i)$ have periodicity $\ell_i^{\rm eff}$.  Compare $A_{xy}$ with \eqref{Axye} and see Section \ref{sec:special} for the origin of the first term in $\phi^{xy}$.

The quantization of $f^x$, $f^y$ and their conjugate variables $\hat f_x$, $\hat f_y$  proceeds in an analogous way as in \cite{paper1}, with the replacement $\ell_x\to \lxeff ,\ell_y\to \lyeff$.  On a lattice, it leads to $N^{L_x^{\rm eff}+L_y^{\rm eff}-1}$ states.

The global considerations constrain the allowed values of $r$.  One way to see that, is to note   we have the operator statement $e^{iN \phi^{xy}}=1$.  Therefore, $r$ should be a multiple of $m/\text{gcd}(N,m)$. This leads to  $\text{gcd}(N,m)$ values.

Combining the above two contributions, we reproduce the ground state degeneracy \eqref{2+1GSD}.

\subsection{Global symmetry operators}\label{sec:2dglobalsymmetry}

Here we compute the ground state degeneracy using the global symmetry operators.
This calculation mirrors the lattice calculation of the ground state degeneracy using the logical operators.

The gauge-invariant local operator $e^{i \phi^{xy}} $ generates a $\bZ_N$ electric global symmetry.
In particular, $e^{i N \phi^{xy}}=1$.
The equation of motion states that
\ie\label{eomphixy}
\partial_x\partial_y\phi^{xy}=0\,.
\fe
The discussion in Section \ref{sec:specialZN} then implies that the $\bZ_N$ electric symmetry can be generated by
\ie
&e^{i\phi^{xy}(x,y_0)},~~x_0\le x<x_0+\lxeff\\
& e^{i\phi^{xy}(x_0,y)},~~y_0\le y<y_0+\lyeff\,,
\fe
for any choice of $(x_0,y_0)$. There is also a $\bZ_{\text{gcd}(N,m)}$ operator
\ie
{\bf V} =  \exp\left[-  i \phi^{xy}(x,y+\lyeff)  + i \phi^{xy}(x,y)\right]  \,.
\fe
The discussion in Section \ref{sec:specialZN} implies that $\bf V$ is independent of $x,y$, despite its appearance.
On a lattice, this leads to $L_x^{\rm eff}+L_y^{\rm eff}-1$  $\bZ_N$ charges and one $\bZ_{\text{gcd}(N,M)}$ charge.

On the other hand, the strip operators
\ie\label{stripoperatorZN}
&\bW_{(x)}  ( x_1,x_2) =  \exp\left[ i \int_{x_1}^{x_2}  dx \oint dy A_{xy}  \right]\,,\\
&\bW_{(y)}  (y_1,y_2) =  \exp\left[ i \oint dx\int_{y_1}^{y_2}  dyA_{xy}  \right]\,
\fe
generate a $\bZ_N$ dipole global symmetry.
They obey the constraint $\bW_{(x)} (0,\lxeff) = \bW_{(y)}(0,\lyeff)$. On a lattice we therefore have $L_x^{\rm eff}+L_y^{\rm eff}-1$ such operators.

There is one more gauge-invariant operator, which is most conveniently expressed in the $\phi-A$ description:
\ie\label{newAxy}
{\bf U} =&
   \exp\Big[ \,
{i} { N\over   \text{gcd}(N,m)} \int_{0}^{\lyeff} dy \int_{ k y \lxeff/\lyeff}^{ky\lxeff/\lyeff+m\lxeff} dx \,\left( \ThetaP(x,0,\lxeff)   - k \ThetaP(y, 0,\lyeff)  \right)  A_{xy} (x,y) \\
&-{i}{ m\over   \text{gcd}(N,m)} \int_{0}^{\lyeff} ds  \,
  \partial_y \phi (  ks \lxeff/\lyeff , s) \,\Big]
\fe
We will motivate this $\bZ_{\text{gcd}(N,m)}$ operator in  Appendix \ref{app:ZN} and discuss its relation to the Wilson strips $\bW_{(i)}$.
One can check that this is indeed a gauge-invariant operator.
Note that the integrand in the first line vanishes in the rectangle $(k-1)\lxeff <x<k \lxeff, 0<y<\lyeff$.

To summarize, just like on an untwisted torus, the $\bZ_N$ theory has a $\bZ_N$ electric and a $\bZ_N$ dipole global symmetry.  Similar to the analysis in \cite{paper1},  there are $L_x^{\rm eff}+L_y^{\rm eff}-1$ $\bZ_N$ operators for each of these symmetries on a lattice.
The novelty on a twisted torus is that there are two additional $\bZ_{\text{gcd}(N,m)}$ symmetries generated by $\bf V$ and $\bf U$.

The $\bZ_{\text{gcd}(N,m)}$ symmetry generated by $\bf U$ in the   $\bZ_N$ theory comes from   the $\bZ_m$ symmetry generated by ${\cal U}(0,0)$ (see \eqref{Zmcharge}) in the $\phi$ theory before Higgsing.
More specifically, when the equation of motion $A_{xy}={1\over N} \partial_x\partial_y\phi$ is imposed, the operator $\bf U$ is equal to ${\cal U}(0,0)^{m/\text{gcd}(N,m)}$  on-shell.

Similarly, the $\bZ_{\text{gcd}(N,m)}$ symmetry generated by $\bf V$ in the   $\bZ_N$ theory comes from   the $\bZ_m$ symmetry generated by $\cal V$ \eqref{Zmchargephiexy} in the $\phi$ theory before Higgsing.
Using the equation of motion  $\partial_x\partial_y\phi^{xy}=0$, we see that ${\bf V}$ is equal to ${\cal V}$ on-shell.

 Let us discuss the commutation relations between these operators.

 In the unitary gauge, where $\phi=0$, \eqref{newAxy} is manifestly an operator in the $\bZ_N$ gauge theory.
 It is then clear that $\bf U$ and $\bf V$ form a $\bZ_{\text{gcd}(N,m)}$ clock and shift algebra.
 This leads to $ \text{gcd}(N,m)$  states in the $\bZ_N$ theory.

The rest of the operators satisfy
\ie\label{nonAbZN}
&e^{i N \phi^{xy}} = \bW_{(x)}^N= \bW_{(y)}^N=1\,,\\
&e^{i  \phi^{xy} (x,y)}  \bW_{(x)} (x_1,x_2)   = e^{2\pi i /N}  \bW_{(x)} (x_1,x_2) e^{i\phi^{xy}(x,y)} \,,~~~~~\text{if } x_1 < x + I\lxeff < x_2 \text{ for some $I\in \bZ$}, \\
&e^{i \phi^{xy} (x,y)}  \bW_{(y)} (y_1,y_2)   = e^{2\pi i /N}  \bW_{(y)} (y_1,y_2) e^{i\phi^{xy}(x,y)} \,,~~~~~\text{if } y_1 < y+I \lyeff < y_2  \text{ for some $I\in \bZ$},
\fe
and they commute otherwise.  (Here, we took for simplicity $x_2-x_1\le\lxeff$ and $y_2-y_1\le \lyeff$.)

We can pick the following basis for them:
\ie\label{ZNcc}
& \bW_{(x)} (x,x+a)\,,~~~~~~~~\exp\left[ i\phi^{xy} ( x, 0)\right]\,,~~~~~~~~~~~~~~~~~~~~~~~~~~~~~~~~~~~(k-1)\lxeff \le x<k\lxeff\,,\\
&\bW_{(y)} (y,y+a)\,,~~~~~~~~\exp\left[ i\phi^{xy}\left  (( k-1) \lxeff,  y\right)- i\phi^{xy} \left(( k-1) \lxeff, 0\right)\right]\,,~~~~~~~0<y<\lyeff\,.
\fe
Here $a$ is an infinitesimal UV regulator, e.g., the lattice spacing.
The range of $x,y$ for these operators is chosen such that they commute with $\bf U,V$.
The pair of operators in each line at the same $x$ or $y$ form a $\bZ_N$ clock and shift algebra,  and they commute otherwise.
On  a lattice, these give $L_x^{\rm eff}+L_y^{\rm eff}-1$ copies of the $\bZ_N$ clock and shift algebra.
This algebra leads to  $N^{\Lxeff+\Lyeff-1}$ states in the $\bZ_N$ theory.

Combining  these two contributions, we reproduce the ground state degeneracy   \eqref{2+1GSD}.

\section{Winding in $3+1$ dimensions}\label{sec:3+1winding}

In this section, we place a classical circle-valued field $\phi$    on a three-dimensional twisted torus.
We perform a twist in the $xy$-plane of the form discussed in Section \ref{sec:GEOMETRY}, but we do not twist in the $xz$-plane or $yz$-plane.
The twist changes the allowed winding configurations relative to those in \cite{paper2} on an untwisted torus.

This discussion is relevant both for the $3+1$-dimensional $\phi$-theory of \cite{paper2} on the twisted torus and for the discussion of the $\bZ_N$ tensor gauge theory in Section \ref{31ZN}.

The winding charges of a circle-valued $\phi$ field  in $3+1$ dimensions are associated with cycles on the $xy$-, $yz$-, and $zx$-planes.
For the $yz$-plane, we will choose the $Y$ cycle and the $Z$ cycle to parameterize these charges, and similarly for the $zx$-plane.
For the $xy$-plane, however, the $X$ and the $Y$ cycles do not generate all the cycles.
Instead, we will choose   the basis $\{X, \tilde X\}$.
The most general possible winding charges are given by (recall the definition \eqref{periodicd})
\ie\label{chargesnb}
&{1\over 2\pi} \oint dx\partial_x\partial_y \phi = \sum_{\beta} W^y_{X\,\beta}\,\deltaP(y, y_\beta,\lyeff)\,,\\
&{1\over 2\pi} \oint dy\partial_x\partial_y \phi = \sum_{\alpha} W^x_{Y\, \alpha}\,\deltaP(x, x_\alpha,\lxeff)\,,\\
&{1\over 2\pi} \oint dz\partial_y\partial_z \phi = \sum_{\beta} W^y_{Z\,\beta}\,\deltaP(y, y_\beta,\lyeff)\,,\\
&{1\over 2\pi} \oint dz\partial_x\partial_z\phi = \sum_{\alpha} W^x_{Z\, \alpha}\,\deltaP(x, x_\alpha,\lxeff)\,,\\
&{1\over 2\pi} \oint dx\partial_z\partial_x \phi = \sum_{\gamma} W^z_{X\, \gamma} \,\deltaP(z, z_\gamma, \ell_z)\,,\\
&{1\over 2\pi} \oint_{\tilde X} (dx\partial_z\partial_x \phi +dy \partial_z\partial_y\phi ) = \sum_{\gamma} W^z_{\tilde X\, \gamma} \deltaP(z, z_\gamma, \ell_z)\,,
\fe
and the $\bZ_m$ charge discussed in Section \ref{sec:transition}.
Here $\{x_\alpha\} , \{y_\beta\},\{z_\gamma\}$ are a finite set of points on the intervals $[0,\lxeff) , \, [0,\lyeff) ,\,[0,\ell_z)$ of the three axes, respectively.
The $W^i_{ I\, \alpha}$'s are integers associated with the points $\{x^i_\alpha\}$ and $I$ labels the cycle.
These charges obey the constraints
\ie\label{Wconstraint2}
&\sum_{\beta} W^y_{X \, \beta} = \sum_{\alpha} W^x_{Y\,\alpha}  \,,\\
&\sum_{\gamma} W^z_{X\,\gamma} =  m \sum_{\alpha}W^x_{Z\,\alpha}\,,\\
&\sum_{\gamma} W^z_{\tilde X\, \gamma} = \sum_{\beta} W^y_{Z\,\beta} +  k \sum_{\alpha} W^x_{Z\,\alpha}\,.
\fe
The winding configurations associated with nontrivial $W^x_{Y\,\alpha} ,W^y_{X\,\beta}$ and the $\bZ_m$ phase have already been discussed in Section \ref{sec:2+1windingphi}.
The rest of the winding configurations are (recall the definition \eqref{ThetaPd})
\ie\label{generalwinding2}
\phi = 2\pi
& \left[\left( {x\over \ell_x} -  k {y\over \ell_y}\right) \sum_{\gamma}  W^z_{ X\, \gamma} \ThetaP(z,z_\gamma,\ell_z)
+{z\over \ell_z} \sum_{\alpha} W^x_{Z\,\alpha} \ThetaP(x, x_\alpha,\lxeff)
\right.\\
& -  {z\over \ell_z}\left( {x\over \ell_x} -  k {y\over \ell_y}\right)  \sum_{\gamma} W^z_{X\,\gamma}  \\
&+\left.{y\over \lyeff} \sum_{\gamma}   W^z_{\tilde X\, \gamma} \ThetaP(z,z_\gamma,\ell_z)
+{z\over \ell_z} \sum_{\beta} W^y_{Z\,\beta}\ThetaP(y, y_\beta,\lyeff)- {z\over \ell_z}{y\over  \lyeff} \sum_{\gamma} W^z_{\tilde X\,\gamma}
\right]
\fe
Here $W^y_{ Z\,\beta} , W^x_{Z\,\beta}, W^z_{X\,\gamma}, W^z_{\tilde X \,\gamma} \in \bZ$ are the integer winding charges obeying  \eqref{Wconstraint2}.

In order to verify that the function $\phi$ of \eqref{generalwinding2} is an allowed configuration, we need to check its periodicity on the torus.
It suffices to check that $\phi$ is a $2\pi$-periodic function along the $Z$, $X$, and $\tilde X$ cycles.
The transition function around the $Z$ cycle is
\ie
g_Z(x,y,z)= \phi(x,y, z+\ell_z) -\phi(x,y, z)
= 2\pi\left[  \sum_{\alpha} W^x_{Z\, \alpha} \ThetaP(x,x_\alpha, \lxeff)    + \sum_\beta W^y_{Z\,\beta} \ThetaP(y,y_\beta,\lyeff) \right]
\fe
The transition function around the $X$ cycle is
\ie
g_X(x,y,z) = \phi(x+m\lxeff, y,z) -\phi(x,y,z)
= 2\pi \sum_\gamma W^z_{X\,\gamma}\ThetaP(z,z_\gamma,\ell_z)\,.
\fe
The transition function around the $\tilde X$ cycle is
\ie
g_{\tilde X}(x,y,z)  = \phi(x+k\lxeff,y+\lyeff,z) -\phi(x,y,z)= 2\pi \sum_\gamma W^z_{\tilde X\,\gamma}\ThetaP(z,z_\gamma,\ell_z)\,.
\fe
Indeed, all these transition functions are $2\pi\bZ$ valued.
Using \eqref{Wconstraint2}, one finds that the cocycle conditions are satisfied.

Combining with the winding configurations in Section \ref{sec:2+1windingphi},  we have   $2(\Lxeff+\Lyeff+L_z)-3$ integer winding charges and one $\bZ_m$ phase   in $3+1$ dimensions on a lattice.

\section{$3+1$-dimensional $\mathbb{Z}_N$ tensor gauge theory}\label{31ZN}

Let us now consider the $\bZ_N$ tensor gauge theory of \cite{paper3}, the continuum limit of the X-cube model  \cite{Vijay:2016phm},  on the twisted torus.  We twist in the $xy$-plane, as in Section \ref{sec:GEOMETRY}, but we do not twist in the $xz$-plane or $yz$-plane.

\subsection{$\phi-A$ theory}\label{sec:3dphiA}

We   start with the Higgs Lagrangian using  $\phi$  and $A$ \cite{paper3}:
\ie\label{Lag1}
{\cal L} =
 - {1 \over 2(2\pi) } \sum_{i\ne j}\hat E^{ij} ( \partial_i \partial_j \phi - NA_{ij})
- {1\over 2\pi }\hat B(\partial_0\phi - NA_0)\,.
\fe
The fields $\hat E^{ij}$ in the $\mathbf{3}'$ and $\hat B$ in the $\mathbf{1}$ serve as Lagrange multipliers. Their coefficients are set such that $\hat E^{ij}$ and $\hat B$ are standardly normalized field strengths of a dual theory.
The $U(1)$ gauge transformation acts on the fields via
\ie\label{Agauge}
&A_0 \sim A_0 +\partial_0\alpha\,,~~~~A_{ij}\sim A_{ij}+\partial_i\partial_j \alpha\,,~~~~
\phi \sim \phi + N\alpha\,,
\fe
with $2\pi$-periodic $\phi$ and $\alpha$, as in Section \ref{sec:3+1winding}.

The equations of motion are given by
\ie\label{eom1}
&\partial_i \partial_j \phi - NA_{ij}=0\,,\\
&\partial_0\phi - NA_0 =0\,,\\
&\hat E^{ij}= \hat B=0\,,
\fe
and they imply the vanishing of the gauge-invariant field strengths of $A$:
\ie\label{EB0}
&E_{ij} = \partial_0 A_{ij} - \partial_i\partial_j  A_0\,,\\
&B_{[ij]k}  =\partial_i A_{jk}-\partial_j A_{ik}\,.
\fe
We will sometimes also use $B_{i(jk)} \equiv B_{[ij]k} + B_{[ik]j}$.

Using the equations of motion \eqref{eom1}, we can solve all the other fields in terms of $\phi$, and the solution space reduces to
\ie
\Big\{ \phi  \Big\}~/~ \phi \sim \phi+N\alpha\,.
\fe
Then, all the $\phi$ configurations can be gauged to a linear combination of the winding modes \eqref{minimalphi}, \eqref{generalwinding2}.  In these linear combinations the coefficients are integers valued in $\{0,...,N-1\}$.

For the purpose of finding the ground state degeneracy, we place the system on a lattice.
From this point on, the analysis of these winding modes is similar to \cite{paper3} if we replace $L_x, L_y$ by $\Lxeff , \Lyeff$.
These winding modes are labeled by $2 \Lxeff +2 \Lyeff +2 L_z -3$ integers valued in $\{0,...,N-1\}$, plus one $\bZ_{\text{gcd}(N,M)}$ phase as in Section \ref{sec:2+1phiA}.
Therefore, the ground state degeneracy is
\ie\label{3dGSD}
N^{2 \Lxeff +2 \Lyeff +2 L_z -3} \text{gcd}(N,M)\,.
\fe

Alternatively, we can compute the ground state degeneracy by choosing a trivialization where the gauge fields are trivial and all the nontrivial information is in the transition functions.
This proceeds along the same lines as in Section \ref{sec:2+1phiA}, and we again arrive at the same result \eqref{3dGSD}.

\subsection{Comments on the $U(1)$ $\hat A$ theory}\label{sec:hatA}

In the previous subsection we  computed the ground state degeneracy of the $\bZ_N$ X-cube model using one of the continuum Lagrangians in \cite{paper3}.
Below we will reproduce the same result using the other dual Lagrangians of the X-cube model in that reference.
These  presentations involve an exotic gauge field  $\hat A$.
Before we discuss these other presentations, we first comment on some new features of the $U(1)$ gauge theory of $\hat A$ on a twisted torus.
We refer the readers to \cite{paper2} for detailed discussion of this gauge theory on an untwisted torus.

The temporal components $\hat A_0^{i(jk)}$ and spatial components $\hat A^{ij}$ are in the $\mathbf{2}$ and $\mathbf{3}'$ of the spatial $S_4$ rotation symmetry.
They are subject to the gauge transformations
\ie \label{hatAgauge}
&\hat A_0^{i(jk)} \sim \hat A_0^{i(jk)}  +\partial_0 \hat \alpha^{i(jk)}\,,\\
&\hat A^{ij} \sim \hat A^{ij}  +\partial_k \hat \alpha^{k(ij)}  \,,\\
\fe
where   $\hat \alpha^{k(ij)}$ is $2\pi$-periodic and transforms in the $\mathbf{2}$ of $S_4$.
The  electric and magnetic fields for $\hat A$ are
\ie\label{hatEB0}
&\hat E^{ij} = \partial_0 \hat A^{ij} - \partial_k \hat A^{k(ij)}_0\,,\\
&\hat B = \frac 12 \sum_{i\neq j}\partial_i\partial_j\hat A^{ij}\,.
\fe

Similar to  Section \ref{sec:transition}, we start with gauge fields on the covering space $\mathbb{R}^3$.
(We limit ourselves to static configurations.)
The identification across a cycle $\cal C$ should involve the circled-valued transition function of the form \eqref{hatAgauge}:
\ie\label{hatAtran}
&\hat A^{ij}  (x+{\cal C}^x, y+{\cal C}^y ,z+{\cal C}^z) = \hat A^{ij}(x,y,z) +\partial_k \hat\gamma^{k(ij)}_{\cal C}(x,y,z)
\fe
where $({\cal C}^x,{\cal C}^y,{\cal C}^z)$ is a vector on a covering space corresponding to the cycle $\cal C$.
Since they transform in the $\mathbf{2}$ of the $S_4$ spatial rotation symmetry, the transition functions are constrained by $e^{ i (\hat\gamma_{\cal C}^{x(yz)}  + \hat \gamma_{\cal C}^{y(zx)} + \hat \gamma_{\cal C}^{z(xy)})}=1$.

Complex matter fields $\hat\Phi^{i(jk)}$ with charge $J$ satisfy $\hat \Phi^{x(yz)}\hat \Phi^{y(zx)}\hat \Phi^{z(xy)}=1$.  They transform under \eqref{hatAgauge} as
\ie\label{Phihatg}
\hat\Phi^{i(jk)} \sim e^{iJ\hat \alpha ^{i(jk)}}\hat\Phi^{i(jk)}
\fe
and under \eqref{hatAtran} as
\ie\label{Phihatgamma}
\hat\Phi^{i(jk)}(x+{\cal C}^x, y+{\cal C}^y ,z+{\cal C}^z) = e^{iJ\hat \gamma_{\cal C} ^{i(jk)}(x,y,z)}\hat\Phi^{i(jk)}(x,y,z)~.
\fe

The composition  of cycles ${\cal C}={\cal A}+{\cal B}$ leads to the cocycle condition:
\ie\label{hatcocycle}
&\exp\left[ i  \hat \gamma_{\cal B}^{i(jk)}  (x,y,z)  + i\hat \gamma_{\cal A}^{i(jk)} (x+{\cal B}^x ,y+{\cal B}^y ,z+{\cal B}^z )\right]\\
&=\exp\left[i  \hat \gamma_{\cal A}^{i(jk)}  (x,y,z)  + i\hat \gamma_{\cal B}^{i(jk)} (x+{\cal A}^x ,y+{\cal A}^y ,z+{\cal A}^z )\right]\,.
\fe

Next, we identify configurations with different transition functions that are related by certain transformations.
Specifically, for any three circle-valued functions $e^{i\hat\beta^{i(jk)}}$ satisfying $e^{ i (\hat\beta^{x(yz)}  + \hat \beta^{y(zx)} + \hat \beta^{z(xy)})}=1$, we identify
\ie\label{hatbeta}
&\exp[i\hat\gamma_{\cal C}^{i(jk)} (x,y,z)] \sim \exp\left[
i \hat \gamma^{i(jk)}_{\cal C} (x,y,z) +i\hat\beta^{i(jk)}(x+{\cal C}^x ,y+{\cal C}^y,z+{\cal C}^z) -i\hat \beta^{i(jk)}(x,y,z)
\right]\,,\\
&\hat A^{ij} (x,y,z) \sim \hat A^{ij}(x,y,z) +\partial_k \hat\beta^{k(ij)}(x,y,z)\\
&\hat\Phi^{i(jk)} (x,y,z)\sim e^{iJ\hat \beta ^{i(jk)}(x,y,z)}\hat\Phi^{i(jk)}(x,y,z)\,.
\fe
If $e^{i\hat\beta^{i(jk)}}$ is single-valued, then this is a gauge transformation and it does not change the transition functions. Otherwise, it relates different trivializations of the same configuration.

The configurations on our three-torus are characterized by the circled-valued transition functions $e^{i \hat\gamma_{\cal C}^{i(jk)}(x,y,z)}$ subject to the cocycle conditions  \eqref{hatcocycle} modulo the identifications \eqref{hatbeta}.

\bigskip\bigskip\centerline {\it Restrict to flat gauge fields}\bigskip

We will be particularly interested in  configurations with  $\hat B=0$.
For such configurations, we can further set locally $\hat A^{ij}=0$  by gauge transformations, and all the information is then contained in the transition functions.

Since $\hat A^{ij}=0$, the transition functions must obey
\ie
\partial_k \hat\gamma_{\cal C}^{k(ij)} =0
\fe
for every $\cal C$.
Using  $e^{ i (\hat\gamma_{\cal C}^{x(yz)}  + \hat \gamma_{\cal C}^{y(zx)} + \hat \gamma_{\cal C}^{z(xy)})}=1$, we find that the transition functions factorize:
\ie\label{hatfactorize}
&e^{i\hat\gamma_{\cal C}^{x(yz)} } = \exp\left[ i f^y_{\cal C}(y) + i f^z_{\cal C}(z)\right]\,,\\
&e^{i\hat\gamma_{\cal C}^{y(zx)} } = \exp\left[ i f^x_{\cal C}(x) - i f^z_{\cal C}(z)\right]\,,\\
&e^{i\hat\gamma_{\cal C}^{z(xy)} } = \exp\left[ -i f^x_{\cal C}(x) - i f^y_{\cal C}(y)\right]\,.
\fe
This parametrization has a zero mode ambiguity
\ie\label{hatzeromode}
f^x_{\cal C}(x) \sim f^x_{\cal C}(x)+c_{\cal C} ,~~~f^y_{\cal C}(y) \sim f^y_{\cal C}(y) -c_{\cal C} ,~~~f^z_{\cal C}(z)\sim f^z_{\cal C}(z)+c_{\cal C}\,.
\fe
We will soon discuss the periodicities of these functions $f^i_{\cal C}(x^i)$.

We will choose $e^{i\hat \gamma^{i(jk)}_X},e^{i\hat\gamma^{i(jk)}_{ \tilde X}},e^{i\hat\gamma^{i(jk)}_ Z}$ associated with the $X,\tilde X, Z$ cycles as our basis for the transition functions, while the others (including the one associated with the $Y$ cycle) are determined in terms of them.  We will use the zero mode ambiguity \eqref{hatzeromode} to set, $e^{if_X^y(0)}= e^{if_{\tilde X}^x(0)}=e^{if_Z^x(0)}=1$.

As discussed above, not all values of $e^{i\hat \gamma^{i(jk)}_{\cal C}}$ are distinct, and we can relate them using $e^{i\hat\beta^{i(jk)}}$.
In order to preserve $\hat A=0$,  $e^{i\hat\beta^{i(jk)}}$ should factorize into three functions of one variable  \ie\label{betafree}
&e^{i\hat\beta^{x(yz)}}=e^{iF^y(y)+iF^z(z)}\,,\\
&e^{i\hat \beta^{y(zx)} } =e^{iF^x(x)-iF^z(z)}\,,\\
&e^{i\hat \beta^{z(xy)}} = e^{-iF^x(x)-iF^y(y)}\,.
\fe
This allows us to set
\ie
e^{if_X^x(x)} =e^{i f_{\tilde X}^y(y)}= e^{if_Z^z(z)}=1\,
\fe
and then the residual freedom in \eqref{betafree} is with functions satisfying
\ie\label{hatresidual}
&e^{iF^x(x+m\lxeff)}  = e^{iF^x(x)}\,,\\
&e^{iF^y(y+\lyeff)}  = e^{iF^y(y)}\,,\\
&e^{iF^z(z+\ell_z)}  = e^{iF^z(z)}\,.
\fe

We are then left with six functions of one variable, $e^{if_X^y(y)}$, $e^{if_X^z(z)}$, $e^{i f_{\tilde X}^x(x)}$, $e^{i f_{\tilde X}^z(z)}$, $e^{i f_Z^x(x)}$, and $e^{i f_Z^y(y)}$, satisfying $e^{if_X^y(0)}= e^{if_{\tilde X}^x(0)}=e^{if_Z^x(0)}=1$.

We now discuss the constraints on these six functions from the cocycle conditions \eqref{hatcocycle} for the $\{X,Z\}$, $\{\tilde X, Z\}$, and $\{X,\tilde X\}$ cycles.
They constrain the functions to satisfy
\ie
&e^{if^y_X(y+\lyeff)} = e^{if^y_X(y)}\,,~~~e^{if^z_X(z+\ell_z)} = e^{if^z_X(z)} \,,~~~e^{if^x_{\tilde X}(x+m\lxeff)} =e^{if^x_{\tilde X}(x)}\,,\\
&e^{if^x_Z(x)+if^y_Z(y)} =e^{ i\tilde f^x_Z(x)+i\tilde f^y_Z(y) + 2\pi i { r\over m} \left( \ThetaP(x,0,\lxeff) - k\ThetaP(y,0,\lyeff) \right)  }\,,\\
&e^{if_{\tilde X}^z(z)}  = e^{- 2\pi i { kr\over m}  \ThetaP(z,0, \ell_z) +i\tilde f_{\tilde X}^z(z) }\,,\\
\fe
where
\ie
&e^{i\tilde f^x_Z(x+\lxeff)}=e^{i\tilde f^x_Z(x)}\,,\\
&e^{i\tilde f^y_Z(y+\lyeff)}=e^{i\tilde f^y_Z(y)}\,,\\
&e^{i\tilde f_{\tilde X}^z(z+\ell_z) } =e^{i\tilde f_{\tilde X}^z(z)}\,.
\fe
Here   $r=0,1,\cdots, m-1$ is an integer that arises from an argument similar to that of Section \ref{sec:special}.

Finally, we can use the residual freedom in $e^{i\hat\beta^{i(jk)}}$ in \eqref{hatresidual} to further restrict the periodicity of $e^{if^x_{\tilde X}(x)}$ to $\lxeff$:
\ie
e^{if^x_{\tilde X}(x+\lxeff) } = e^{if^x_{\tilde X}(x)}\,.
\fe

To summarize, the $\hat B=0$ configurations can be described by $\hat A=0$ and transition functions that are characterized by six circle-valued functions of one variable $e^{if_X^y(y)}$, $ e^{if_X^z(z)}$, $e^{i f_{\tilde X}^x(x)}$, $e^{i\tilde f_{\tilde X}^z(z)}$, $e^{i\tilde f_Z^x(x)}$, and $e^{i\tilde f_Z^y(y)}$, satisfying $e^{if_X^y(0)}= e^{if_{\tilde X}^x(0)}=e^{if_Z^x(0)}=1$, and with periodicities $\lxeff$, $ \lyeff$, or $\ell_z$, as well as a $\bZ_m$-valued integer $r$.
These functions are physical and they contribute to the holonomies.

On a lattice, these lead to $2(L_x^{\rm eff}+L_y^{\rm eff}+L_z)-3$ distinct $U(1)$ phases and one $\bZ_M$-valued integer.
(Recall that we label the lattice quantities by upper case letters and their continuum counterparts by lower case letters.)

The main novelty on a twisted torus is the $m$  configurations labeled by $r$.
Quantum mechanically, most of these configurations acquire infinite energies \cite{paper2}.
However, these  $m$ degenerate states remain zero-energy states.

Similar to the discussion in Section \ref{sec:commentphi}, the $3+1$-dimensional $\phi$-theory also has  $m$ degenerate ground states.
This provides another check of the duality  derived in \cite{paper2} between the pure $U(1)$ gauge theory of $\hat A$ and the  $\phi$-theory.

\subsection{$\hat \phi-\hat A$ theory}\label{sec:hatphiA}

We now proceed to compute the number of ground states from the perspective of a different Higgs Lagrangian using a circle-valued field  $\hat\phi^{i(jk)}$ in the $\mathbf{2}$ of $S_4$  and gauge fields $(\hat A_0^{i(jk)} , \hat A^{ij})$ in the $(\mathbf{2},\mathbf{3}')$ of $S_4$ \cite{paper2,paper3}. In comparing with the discussion around \eqref{Phihatg}, $\hat \Phi^{i(jk)}=e^{i\hat\phi^{i(jk)}}$ with charge $J=N$.  These fields are subject to the gauge transformation
\ie\label{hatgauge}
&\hat A_0^{i(jk)} \sim \hat A_0^{i(jk)}  +\partial_0 \hat \alpha^{i(jk)}\,,\\
&\hat A^{ij} \sim \hat A^{ij}  +\partial_k \hat \alpha^{k(ij)}  \,,\\
&\hat \phi^{k(ij)} \sim \hat \phi^{k(ij)}+N\hat \alpha^{k(ij)}
\fe
where $\hat\phi^{k(ij)}$ and $\hat \alpha^{k(ij)}$ are $2\pi$-periodic and transform in the $\mathbf{2}$ of $S_4$, as in Section \ref{sec:3+1winding}.
The Lagrangian is:
\ie
{\cal L}  = {1\over 2(2\pi)} \sum_{i\neq j} E_{ij}\left (\partial_k \hat\phi^{k(ij)} -N\hat A^{ij}\right)   - {1\over 6(2\pi)} \sum_{i\neq j\neq k} B_{k(ij)} \left(\partial_0\hat\phi^{k(ij)} -N \hat A^{k(ij)}_0\right) \,.
\fe

We can choose the unitary gauge $e^{i\hat\phi^{i(jk)}}=1$, and then the equation of motion sets the $\hat B=0$.
Following the discussion in Section \ref{sec:hatA}, we can choose $\hat A=0$, and then all the information is contained in the transition functions $e^{i\hat\gamma_{\cal C}^{i(jk)}}$. In order to preserve the gauge choice $e^{i\hat\phi^{i(jk)}}=1$, they should be $\bZ_N$ phases.
These transition functions are parameterized by six $\bZ_N$-valued functions  $e^{if_X^y(y)}$, $e^{if_X^z(z)}$, $e^{i f_{\tilde X}^x(x)}$, $e^{i\tilde f_{\tilde X}^z(z)}$, $e^{i\tilde f_Z^x(x)}$, and $e^{i\tilde f_Z^y(y)}$, satisfying $e^{if_X^y(0)}= e^{if_{\tilde X}^x(0)}=e^{if_Z^x(0)}=1$ with periodicities $\lxeff, \lyeff$, or $\ell_z$, together with a $\bZ_{\text{gcd}(N,m)}$-valued integer.
On a lattice, this leads to the ground state degeneracy  \eqref{3dGSD}.

Of special importance are the $\text{gcd}(N,m)$ states characterized by
\ie\label{gcdstates}
&e^{i\hat\phi^{i(jk)}} =1,~~\hat A_0^{i(jk)}=\hat A^{ij}=0\,,~~~e^{i\hat\gamma_X^{i(jk)}}=1\,,\\
&e^{i\hat \gamma_Z^{x(yz)}}= e^{ {2\pi\over \text{gcd}(N,m)}  k r\ThetaP(y,0,\lyeff) }\,,~~~~
e^{i\hat \gamma_Z^{y(zx)}}= e^{- {2\pi\over \text{gcd}(N,m)} r \ThetaP(x,0,\lxeff)}\,,~~~~\\
&e^{i\hat \gamma_Z^{z(xy)}} =e^{ {2\pi\over \text{gcd}(N,m)}  r (\ThetaP(x,0,\lxeff) -k\ThetaP(y,0,\lyeff))}\,,\\
&e^{i\hat \gamma_{\tilde X}^{x(yz)}} = e^{-i\hat\gamma_{\tilde X}^{y(zx)}}=  e^{ {2\pi \over \text{gcd}(N,m)} kr \ThetaP(z,0,\ell_z) }\,,~~~~
e^{i\hat\gamma_{\tilde X}^{z(xy)}}=1\,,
\fe
where $r=0,1,\cdots,\text{gcd}(N,m)-1$.

\subsection{$BF$-type $\bZ_N$ tensor gauge theory}\label{sec:ZN}

Now we consider a dual presentation of the $\bZ_N$ tensor gauge theory \cite{Slagle:2017wrc,paper3}, which will permit a different computation of the ground state degeneracy. This presentation involves gauge fields $(A_0 ,A_{ij})$ in the $(\mathbf{1}, \mathbf{3}')$ of $S_4$, and $(\hat A_0^{i(jk)}, \hat A^{ij})$ in the $(\mathbf{2}, \mathbf{3}')$ of $S_4$.
These fields are subject to the gauge transformations \eqref{Agauge} and \eqref{hatAgauge}, and their field strengths are \eqref{EB0} and \eqref{hatEB0}.

The $BF$-type Lagrangian in this presentation is:
\ie\label{BFLag1}
{\cal L} = {N\over 2\pi }  \left(
\frac 12 \sum_{i\ne j} A_{ij} \hat E^{ij}   + A_0 \hat B
\right) \,.
\fe
As in \cite{paper3}, we work in temporal gauge, setting $A_0=0$ and $\hat A_0^{i(jk)}=0$.

The analysis of the terms involving $A_{xy},\hat A^{xy}$ proceeds in a similar way as in Section \ref{sec:2+1BF}.
The quantization of the electric modes for $A_{xy},\hat A^{xy}$ leads to  the bulk part of the spectrum, which becomes $N^{L_x^{\rm eff}+L_y^{\rm eff}-1}$ states on a lattice.
In addition, there are $\text{gcd}(N,m)$ states \eqref{gcdstates} coming from   the transition functions of the $\hat A$ gauge theory.
On a lattice, we have in total $N^{L_x^{\rm eff}+L_y^{\rm eff}-1}\text{gcd}(N,M)$  states.
We will henceforth focus on the modes associated with $A_{zx},A_{yz},\hat A^{zx},\hat A^{yz}$.

The solutions to the Gauss law constraints modulo gauge transformations are
\ie\label{AAhat}
&A_{zx}  = {1\over \ell_x }  f^z_{zx}(t,z)  +{1\over \ell_z  }  f^x _{zx}(t,x)  \, ,\\
&A_{yz}  = {1\over \ell_z  }  f^y _{yz}(t,y) +{1\over \ell_y }  f^z_{yz}(t,z)   \, ,\\
&\hat A^{zx}  = {1\over \ell_y } \hat f^{zx}_z (t,z ) + {1\over \ell_y } \hat f^{zx}_x(t,x) \,,\\
&\hat A^{yz}  = {1\over \ell_x } \hat f^{yz}_x (t,x ) + {1\over \ell_x } \hat f^{yz}_z(t,z) \, .
\fe
These functions are periodic in $x\sim x+\lxeff$, $y\sim y+\lyeff$, and $z\sim z+\ell_z$.

The functions $f$ and $\hat f$ are subject to a redundancy due to the  zero modes \cite{paper3}.
To remove the redundancy of $f$, we define the following  combinations:
\ie
\bar f^i_{ij} (t,x^i) = f^i_{ij}(t,x^i) +{1\over \ell_i } \oint dx^j f^j_{ij} (t,x^j)\,.
\fe
They are subject to the constraint
\ie\label{barfconstraint}
\oint dx^i \bar{f}^i_{ij}(t, x^i) = \oint dx^j \bar{f}^j_{ij}(t, x^j) \,.
\fe
We will use these constraints to solve for the modes $\bar f^z_{zx}(t,z=0), \bar f^z_{yz}(t,z=0)$  in terms of the others.
Correspondingly, we use the  redundancy to set their conjugate variables $\hat f_z^{zx}(t,z=0), \hat f_z^{yz}(t,z=0)$   to zero.

Let us now discuss the periodicities of the modes $f$.
Using the gauge transformations $\alpha$ of the form  \eqref{minimalphi}, \eqref{generalwinding2}, we find that different components of $\bar f^i_{ij}$ have correlated, delta function periodicities.
To diagonalize these periodicities, we define
\ie
&\bar f_{\tilde X}(t,z) ={1\over m}\bar f^z_{yz}(t,z)+  {k \over m}\bar f^z_{zx} (t,z)\,.
\fe
Then the modes  $\bar f^z_{zx}(t,z), \bar f_{\tilde X}(t,z) , \bar f^x_{zx}(t,x),\bar f^y_{yz}(t,y)$ have independent periodicities. For example,
\ie
\bar f_{\tilde X}(t,z) \sim \bar f_{\tilde X}(t,z)+2\pi \deltaP(z, z_0, \ell_z)\,,~~~~z\ne 0~{\rm mod} ~\ell_z
\fe
for each $z_0$. The other three modes have similar delta function periodicities.

We now turn to the periodicities of the modes $\hat f$.
Their periodicities  arise from the large  gauge transformations:
\ie\label{hatgauge}
&{1\over 2\pi} \hat\alpha^{x(yz)}  =  \left( {x\over \ell_x}- \langle \tilde X,Y\rangle {y\over \ell_y} \right) W^y_x(y)  + {x\over \ell_x}W^z_x(z) -  {y\over \ell_y}\, W^z_y(z) -  {z\over \ell_z}\, W^y_z(y),\\
&{1\over 2\pi}\hat\alpha^{y(zx)}  =  {y\over \ell_y}W^z_y(z) +\left( { y\over \ell_y}- \langle \tilde Y,X\rangle {x\over \ell_x } \right) W^x_y(x)- {z\over\ell_z} \, W^x_z(x) -  {x\over\ell_x} \,W^z_x(z)\,,\\
&{1\over 2\pi}\hat\alpha^{z(xy)}  = {z\over \ell_z} \left[ W^x_z(x) +W^y_z(y) \right]  -  \left( { y\over \ell_y}- \langle \tilde Y,X\rangle {x\over \ell_x } \right)  \, W^x_y(x) -  \left( {x\over \ell_x}- \langle \tilde X,Y\rangle {y\over \ell_y} \right) \,W^y_x(y)\,.
\fe
where $W^i_j(x^i)\in \bZ$.
We will motivate these gauge transformations in Appendix \ref{sec:hatwinding} where we discuss the winding configurations of a classical field in the $\mathbf{2}$ on our twisted torus.

These gauge transformations correlate the integer-valued periodicities of different components of $\hat f_i^{ij}$.
To diagonalize these periodicities, we define
\ie
 &{\hat f}^{\tilde X} (t,z)\equiv \frac{1}{m} {\hat f}^{zx}_z(t,z) - \frac{k}{m} \hat f^{yz}_z(t,z)\,.
\fe
Then the modes  ${\hat f}^{yz}_z(t,z)  , {\hat f}^{\tilde X}(t,z) ,{\hat f}^{zx}_x(t,x),{\hat f}^{yz}_y(t,y) $ have  independent, pointwise $2\pi \bZ$ periodicities.

Written in this basis of $\bar f$ and $\hat f$ with independent periodicities, the Lagrangian is diagonalized:
\ie
&  {N\over 2\pi} \int_0^{\lxeff} dx \hat f^{zx}_x (t,x) \partial_0 \bar f^x_{zx}(t,x)   + {N\over 2\pi} \int_0^{\lyeff} dy \hat f^{yz}_y (t,y) \partial_0 \bar f^y_{yz}(t,y)\\
&+{N\over 2\pi}\int_{0^+}^{\ell_z} dz  \left( \hat f^{\tilde X}(t,z)  \partial_0 \bar f^z_{zx}(t,z) +   \hat f^{yz}_z (t,z) \partial_0 \bar f_{\tilde X}(t,z)  \right)  \,.
\fe
Recall that we have removed the modes at $z=0$ using the constraint \eqref{barfconstraint} of $f$ and the  redundancy  of $\hat f$.
Quantizing these modes on a lattice, we obtain a Hilbert space with $N^{L_x^{\rm eff}+L_y^{\rm eff}+2L_z-2}$ zero-energy states.
Combining with the contributions from $A_{xy},\hat A^{xy}$, we  reproduce the ground state degeneracy   in \eqref{3dGSD}.

\subsection{Global symmetry operators}

We can compute the ground state degeneracy in yet another way, using the $\mathbb{Z}_N$ symmetry operators of the theory.  As we will see, the twist leads to various novelties that are not present when the system is placed on the untwisted torus.  Even though, as in \cite{paper3}, we will use continuum notation, this approach is easily related to the corresponding lattice analysis using logical operators. In particular, it mirrors the recent lattice discussion in \cite{1821601}.

Let us start with the global symmetry operators depending on  $A_{xy}$ and its conjugate variable $\hat A^{xy}$.
This part of the analysis proceeds as in Section \ref{sec:2dglobalsymmetry}, with $e^{i\phi^{xy}}$ there replaced by $e^{i \oint dz \hat A^{xy}}$.  We find  $N^{L_x^{\rm eff}+L_y^{\rm eff}-1}\text{gcd}(N,M)$ states.

We now turn to the remaining gauge-invariant operators.
The operators built of  $A_{zx},A_{yz}$ are the Wilson strips:
\ie\label{Wxy}
&\bW_{(x)} (x_1,x_2, Z )= \exp\left[ i  \int_{x_1}^{x_2}dx \oint dzA_{zx}\right]\,,\\
&\bW_{(y)} (y_1,y_2, Z )= \exp\left[ i  \int_{y_1}^{y_2}dy \oint dzA_{yz}\right]\,.
\fe
and
\ie\label{Wz}
&\bW_{(z)}(z_1,z_2,X)=  \exp\left[ i   \int_{z_1}^{z_2} dz \oint dx A_{zx}   \right] \,,\\
&\bW_{(z)}(z_1,z_2,{\tilde X})=  \exp\left[ i   \int_{z_1}^{z_2} dz\oint_{\tilde X}\left(  dx \, A_{zx} + dy\,  A_{yz}  \right) \right] \,.
\fe
(Here, for simplicity, we took $0<x_2-x_1\le \lxeff$, $0< y_2-y_1\le \lyeff$, and $0<z_2-z_1\le \ell_z$.)
More generally, we can study Wilson strip operators $\bW_{(z)}(z_1,z_2,{\cal C})$ associated with any curve   ${\cal C}$ on the twisted $xy$-torus.  They depend only on the homology class of $\cal C$ and not on the explicit representative.
Since $\{X,{\tilde X}\}$ form a basis of $\Gamma= H_1(T^2,\bZ)$, every Wilson strip $\bW_{(z)}(z_1,z_2,{\cal C})$  can be generated by \eqref{Wz}.

The Wilson strips are subject to two constraints, which come from viewing the same integral in two different ways:
\ie\label{Wilsonconstraint}
&\bW_{(x)}(0,\lxeff,Z) ^m =  \bW_{(z)} (0,\ell_z,X)\,,\\
&\bW_{(y)}(0,\lyeff,Z)\bW_{(x)}(0,\lxeff,Z) ^{k} =  \bW_{(z)} (0,\ell_z,\tilde X)\,.
\fe

Next, the gauge-invariant operators of $\hat A^{zx},\hat A^{yz}$ involve the Wilson lines along the $X$ and the $Y$ cycles:
\ie\label{hatWilson}
&\hat \bW^x (y,z) =\exp\left[ i\oint dx \hat A^{yz} \right]\,,\\
&\hat \bW^y(x,z) = \exp\left[ i \oint dy \hat A^{zx}\right]\,.
\fe

The vanishing of the magnetic field of $\hat A$ implies that the $\bZ_N$ symmetry operator $\hat \bW^x(y,z)$  factorizes \cite{paper3}:
\ie\label{factorize}
\hat \bW^x( y,z) = \hat \bW^x_y(y) \hat \bW^x_z(z)
\fe and similarly in the other directions

There is one important novelty here, which was not present in the case of the untwisted torus: the Wilson line operators $\hat \bW^x,\hat\bW^y,\hat\bW^z$ do not generate all the gauge invariant operators constructed out of $\hat A$.  Consider the Wilson strip operators of $\hat A$ \cite{paper3}:\footnote{Both the Wilson lines \eqref{hatWilson} and the Wilson strip \eqref{hatP} can be extended in time to become defects in the $\bZ_N$ theory. They are the continuum representations of the probe limits of a single lineon and a dipole of lineons (which is a planon) of the X-cube model, respectively. See \cite{paper3} for more details.}
\ie\label{hatP}
\hat \bP(z_1,z_2 , {\cal C} )  = \exp\left[
i \int _{z_1}^{z_2} dz \oint_{\cal C}\left( \partial_z \hat A^{yz}dx - \partial_z \hat A^{zx} dy-\partial_y \hat A^{xy}dy\right)
\right]\,.
\fe
Since the magnetic field of $\hat A$ vanishes, this strip operator $\hat \bP$ depends only on the homology class $[\cal C]$ of the curve $\cal C$ on the twisted $xy$-torus, and not on its representative.

Let us first consider the case when $[\cal C]$ is a cycle of $\hat\Gamma$, i.e., the sublattice of $\Gamma$ generated by the $X$ and $Y$ cycles. That is, $[{\cal C}]  = n_xX+n_yY$ with $n_x,n_y\in \bZ$.
In this case we can choose the representative $\cal C$ in \eqref{hatP} to  first go around the $X$ cycle  $n_x$ times, and then around the $Y$ cycle $n_y$ times.
For this choice of $\cal C$, the term $\partial_y\hat A^{xy}$ in \eqref{hatP} does not contribute to the integral, and the strip operator can be written in terms of the Wilson lines:
\ie
\hat \bP (z_1,z_2 ,{\cal C})  = \hat \bW^x (y, z_2)^{n_x} \hat \bW^y (x,z_2)^{-n_y}\hat \bW^x (y, z_1) ^{-n_x}\hat \bW^y (x,z_1)^{n_y}\,.
\fe
Note that the negative sign in the exponent in \eqref{hatP} leads to negative signs in the exponents here.  It is important that because of \eqref{factorize} this $\hat \bP(z_1,z_2,{\cal C})$ is independent of $x$ and $y$.

However, if $[{\cal C}]$ is not a cycle in $\hat\Gamma$, then the strip operator $\hat \bP(z_1,z_2,{\cal C})$ is \textit{not} generated by the Wilson lines.
We should then include these operators as independent $\bZ_N$ operators in addition to  \eqref{hatWilson}.
Note that on an untwisted torus, $\hat \Gamma=\Gamma$ and therefore it suffices to study the Wilson lines $\hat \bW^x,\hat\bW^y,\hat\bW^z$.

We conclude that the gauge-invariant operator built out of $A_{zx},A_{yz},\hat A^{zx},\hat A^{yz}$ are generated by  \eqref{Wxy}, \eqref{Wz}, \eqref{hatWilson}, and \eqref{hatP}.
We can group these operators as follows:
\ie\label{pairs}
&\bW_{(x)}(x,x+a , Z)\,,~~~~\hat \bW^y(x,z=0)\,,~~~~~\, 0\le x<\lxeff\,,\\
&\bW_{(y)}(y,y+a , Z)\,,~~~~\hat \bW^x(y,z=0)\,,~~~~~\, 0\le y<\lyeff\,,\\
&\bW_{(z)}(z,z+a , \tilde X)\,,~~~~\hat \bP(0,z,  X)\,,~~~~~~~~~0 <z<\ell_z\,,\\
&\bW_{(z)}(z,z+a ,  X)\,,~~~~\hat \bP(0,z,\tilde  X)^{-1}\,,~~~~~~0 <z<\ell_z\,.
\fe
Here $a$ is an infinitesimal UV regulator, e.g., lattice spacing.  Using \eqref{Wilsonconstraint}, the operators $\bW_{(z)}(z,z+a , \tilde X)$ and $\bW_{(z)}(z,z+a , \tilde X)$ at $z=0$ can be solved in terms of $\bW_{(x)}$ and $\bW_{(y)}$ and therefore they are not included as independent operators in \eqref{pairs}.
 The operators $\hat \bW^x$ and $\hat \bW^y$ at $z>0$ can be generated by those at $z=0$ and $\hat \bP(0,z,X),\hat \bP(0,z,\tilde X)^{-1}$.

The pair of operators at the same point in space in each line in \eqref{pairs} form a $\bZ_N$ clock and shift algebra, and operators at different points  in space or on different lines in \eqref{pairs} commute with each other.
On a lattice, this gives rise to $L_x^{\rm eff}+L_y^{\rm eff}+2L_z-2$ copies of the $\bZ_N$ clock and shift algebra.
The dimension of the minimal representation of this algebra is $N^{L_x^{\rm eff}+L_y^{\rm eff}+2L_z-2}$.
Combining with the contributions from the operators of $A_{xy},\hat A^{xy}$, we have reproduced the ground state degeneracy \eqref{3dGSD}.

\section*{Acknowledgements}

We thank M.~Hermele for sharing \cite{1821601} with us long before its publication and for answering many questions.  We also thank M.~Cheng for sharing the results of the unpublished work \cite{Meng} and for crucial discussions about them.
We are especially grateful to P.~Gorantla and H.T.~Lam for their critical reading of the manuscript and for helpful conversations.
The work of TR at the Institute for Advanced Study was supported by the Roger Dashen Membership and by
NSF grant PHY-191129. The work of TR at the University of California, Berkeley, was supported by NSF grant  PHY1820912, the Simons Foundation, and the Berkeley Center for Theoretical Physics.
The work of NS was supported in part by DOE grant DE$-$SC0009988.  NS and SHS were also supported by the Simons Collaboration on Ultra-Quantum Matter, which is a grant from the Simons Foundation (651440, NS).
SHS  thanks the Department  of Physics at National Taiwan University for its hospitality during the course of this work.
Opinions and conclusions expressed here are those of the authors and do not necessarily reflect the views of funding agencies.

\appendix

\section{$2+1$ dimensional plaquette Ising model}\label{app:2+1lattice}

In this appendix, we compute the ground state degeneracy of the $2+1$-dimensional $\bZ_N$ plaquette Ising model on our twisted lattice.
We will assume the absence of the transverse field and that we are in the broken phase. The low energy limit of this lattice model  is the $\bZ_N$ tensor gauge theory of \cite{paper1}.

We will analyze the model in the Hamiltonian formalism.
On every site $(\hat x,\hat y)$ there is a pair of $\bZ_N$ clock and shift operators $U(\hat x,\hat y), V(\hat x,\hat y)$ that obey $V(\hat x,\hat y) U(\hat x,\hat y) = e^{2\pi i /N} U(\hat x,\hat y) V(\hat x,\hat y)$ and $U(\hat x,\hat y)^N = V(\hat x,\hat y)^N=1$.
The Hamiltonian is
\ie
H   =  - \beta \sum_{(\hat x,\hat y)} V(\hat x,\hat y) V(\hat x+1,\hat y)^{-1} V(\hat x,\hat y+1)^{-1} V(\hat x+1,\hat y+1) +c.c.\,.
\fe
Since there is no $U$ in the Hamiltonian, we can diagonalize the Hilbert space at every site using $V$.

Let the eigenvalue of $V(\hat x,\hat y)$ be $s(\hat x,\hat y)$, which is also a $\bZ_N$ phase, i.e., $s(\hat x,\hat y)^N=1$.
The translations along the $X$ and $\tilde X$ cycles imply that (see \eqref{periodicityLX})
\ie\label{stranslations}
&X:~s \left (\hat x+M L_x^{\rm eff} , \hat y \right )  =s(\hat x,\hat y)\,,\\
&\tilde X:~s\left (\hat x+ K L_x^{\rm eff} \,,\, \hat y+ L_y^{\rm eff}\right)  =s(\hat x,\hat y)\,.
\fe

Let us find the ground states.  We need to find
$\{ s(\hat x,\hat y)\}$ subject to the constraint:
\ie\label{sconstraint}
s(\hat x,\hat y) \, s(\hat x+1,\hat y)^{-1} \, s(\hat x,\hat y+1)^{-1} \,s (\hat x+1,\hat y+1)=1\,,
\fe
for all lattice sites $(\hat x, \hat y)$.
This is a lattice version of the analysis of $e^{i\phi^{xy}}$  in Section \ref{sec:2dglobalsymmetry}.
We will follow  steps similar to the steps there and will reproduce the answer in the continuum \eqref{2+1GSD}.

Locally, \eqref{sconstraint} is solved by
\ie
s(\hat x,\hat y)= s(1,1)^{-1}s(\hat x,1)s(1,\hat y) ~,
\fe
thus reducing the number of independent variables to $L_x+\Lyeff-1$.  Next, we impose the boundary conditions  \eqref{stranslations}:
\ie
&s \left(\hat x+M L_x^{\rm eff} ,1 \right)  =s (\hat x,1 )\,,\\
&s \left (\hat x+ K L_x^{\rm eff} \,,\, 1\right)s\left (1\,,\, \hat y+ L_y^{\rm eff}\right)  =s(\hat x,1)s(1,\hat y)~.
\fe
This leads to
\ie\label{scriptN}
&s(\hat x+\Lxeff,1)=\eta s(\hat x,1) \\
&s(1, \hat y +\Lyeff) = \eta^{- K }s(1,\hat y)
\fe
with a constant $\eta$ satisfying $\eta^M=\eta^N=1$ and therefore $\eta^{\gcd(N,M)}=1$.

We conclude that the independent solutions are labeled by $\Lxeff+\Lyeff-1$ integers modulo $N$ from $s(\hat x,1)$ and $s(1,\hat y) $ and an integer modulo $\gcd(N,M)$ from $\eta$.  So we end up with
\ie
N^{\Lxeff+\Lyeff -1}\gcd(N,M)
\fe
solutions.

\section{Invariants of the transition functions for $\phi$ on a two-torus}\label{app:counting}

In this appendix, we discuss the invariants for the transition functions $g_{\cal C}(x,y)$ under the identification \eqref{nonperiodic}.
Our goal is to show that all the invariants are given by the $U(1)$ charges $Q^x(x), Q^y(y)$ in \eqref{U1charges} (subject to \eqref{nxy}), and one integer modulo $m$ charge \eqref{Zmcharge}.

Starting with a generic  $\phi$ configuration, we can always subtract from it a standard configuration with the same  $U(1)$ charges.  The resulting $\phi$ configuration has vanishing
$U(1)$ charges.  Therefore, it is enough to consider such configurations.  We are going to show that for them the only remaining invariant is a $\bZ_m$ charge.  This means that we start with a configuration with
\ie
&g_X  = 2\pi n^x_X(x)\,,\\
&g_Y= 2\pi n^y_Y(y)\,,
\fe
and $n_{xy}=0$.
Since there is also  invariant information in $g_{\tilde X}$, we should take it into account.
\eqref{gY}  leads to
\ie\label{gY0}
&n^y_Y(y)  = \sum_{I=0}^{m-1}   n_{\tilde X}^y(y+I\lyeff)   -{\cal N}\,,\\
&0=  {\cal N} - \sum_{J=1}^k n_X^x(x-Jm\lxeff)+\sum_{I=1}^m n^x_{\tilde X}(x-I k \lxeff)\,,
\fe
for some integer $\cal N$.
The cocycle condition  \eqref{XtXcocycle} leads to
\ie\label{cocycle0}
n^x_{\tilde X}(x+m\lxeff)  - n^x_{\tilde X} (x) = n^x_X (x+k\lxeff) - n^x_X(x)\,.
\fe
  The freedom in splitting the zero mode between $n_{\tilde X}^x(x)$ and $n_{\tilde X}^y(y)$ leads to the following identification
\ie\label{scriptN}
&n^x_{\tilde X}(x) \sim n^x_{\tilde X}(x) + N _{\tilde X}\,,\\
&n^y_{\tilde X}(y)\sim n^y_{\tilde X}(y)  - N_{\tilde X}\,,\\
&{\cal N}\sim {\cal N} -mN _{\tilde X}\,.
\fe
Therefore, only 
\ie
&\exp(2\pi i {\cal N}/m) = \exp\left[ -{2\pi i \over m}\sum_{I=1}^m n^x_{\tilde X}(x-I k \lxeff)+  {2\pi i \over m}\sum_{J=1}^k n_X^x(x-Jm\lxeff) \right]\\
&=\exp\left[ -{2\pi i \over m}\sum_{I=0}^{m-1} n^x_{\tilde X}(x+I  \lxeff)+  {2\pi i \over m}\sum_{J=0}^{k-1} n_X^x(x+J\lxeff) \right]
\fe
is meaningful.  
In the second line, we have used \eqref{cocycle0}. 
Indeed,  this agrees with the $\bZ_m$ charge in \eqref{Zmcharge} in the special case when all the $U(1)$ charges $Q^x(x), Q^y(y)$ vanish.

To complete the counting, we use the same strategy as above.  We subtract from our configuration a standard configuration with the same nonzero $\bZ_m$ charge and find a configuration with vanishing $\bZ_m$ charge.  We are going to show that in this case there is no other invariant information.

Using \eqref{nonperiodic}, we can choose   $n_{\tilde X}^x(x) = n_{\tilde X}^y(y)=0$.
Then \eqref{gY0}  implies $n^y_Y(y)=0$ and
\ie\label{sumn}
0=  \sum_{J=1}^k n^x_X (x-Jm\lxeff)\,.
\fe
The only remaining transition function  $n^x_X(x)$ has periodicity $n^x_X(x+k\lxeff) = n^x_X(x)$
and is subject to a residual  identification:
\ie
&n^x_X(x) \sim n^x_X(x) +  n^x( x+ m\lxeff) -n^x(x)\,,
\fe
where  $n^x(x)$ satisfies
\ie
&n^x(x+k\lxeff) = n^x(x)\,.
\fe

Finally, we show that the remaining transition function $n^x_X(x)$ can be set to zero as follows.
Since gcd$(m,k)=1$, we can parameterize every point $x$ as $x= \tilde x -R m\lxeff + P k\lxeff$ for some $0\le \tilde x <\lxeff$ and $R,P\in \bZ$.
This parametrization of $x$ in terms of $(\tilde x, R,P)$ has the ambiguity $(\tilde x, R,P)\sim (\tilde x, R+k,P-m)$.
Using this parametrization, we choose $n^x(x)$ to be
\ie
n^x(x)  =-  \sum_{I=1}^R n^x_X(\tilde x-Im\lxeff)\,.
\fe
The condition \eqref{sumn} ensures that this $n^x(x)$ is invariant under the above ambiguity and has periodicity $k\lxeff$.
This residual identification then removes all the remaining transition functions.

To conclude, we have shown that all the  invariant information in the transition functions is captured by $Q^x(x), Q^y(y)$   in \eqref{U1charges} and one $\bZ_m$ charge \eqref{Zmcharge}.

\section{Additional operators in the $\phi$-theory }\label{app:Uphi0}

In this appendix, we discuss some additional operators in the  $\phi$ theory.
These include the $U(1)$ charges $Q^x(x), Q^y(y)$ \eqref{U1charges} and the $\bZ_m$ charge \eqref{Zmcharge}.

We start with a first attempt.  We want to use the local winding current   ${1\over 2\pi} \partial_x\partial_y\phi$ to construct an operator by integrating it  against a certain profile function $\phi_0(x,y)$:
\ie\label{firstattempt}
``\exp\left[ {i\over2\pi} \int _{T^2}  \phi_0(x,y)   \partial_x\partial_y\phi(x,y)\right]"
\fe
Here  $\phi_0(x,y)$ is another classical background configuration, which is distinct from our field $\phi(x,y)$. (For the purpose of this discussion, our field $\phi(x,y)$ is also a classical field.)
Both $\phi_0$ and $\phi$ obey the rules in Section \ref{sec:2+1winding}.

This definition, however, is not precise.
First,  since both $\phi_0$ and $\phi$ are not real-valued functions on the torus, the integral generally depends on the choice of the fundamental domain for the  torus.
Second, this expression is not invariant under the gauge transformation \eqref{ngauge} for $\phi_0$.

In the rest of this appendix, we will give a precise definition of this operator that does not suffer from the issues above.
See \cite{Cordova:2019jnf} for a closely related discussion in other more familiar models.

For simplicity, we will set $\lxeff=\lyeff=1$ in this appendix.

We claim that the more precise version of the operator \eqref{firstattempt} is
\ie\label{Uphi0}
&{\cal U}(\phi_0)=
   \exp\Big[
{i\over 2\pi} \int_{y_*}^{y_*+1} dy \int_{\tilde x_*+ k y}^{\tilde x_*+ky+m} dx \, \phi_0(x,y) \partial_x\partial_y \phi(x,y) \\
&  - {i\over 2\pi}  \int_{\tilde x_* + ky_*} ^{\tilde x_* +ky_*+m} dx \, g_{\tilde X}^0(x, y_*) \partial_x \phi(x,y_* )\\
&-{i\over2\pi} \int_{y_*}^{y_*+1} ds  \,
\left[ g_X^0( \tilde x_* + ks  , s)   \partial_y \phi ( \tilde x_* + ks , s)+k\partial_x g_X^0( \tilde x_* + ks  )  \,\phi ( \tilde x_* + ks  , s)\right]\\
&+i n^0_{x\tilde x} \, \phi(\tilde x_* + ky_* ,y_*) \Big]
\fe
where $\tilde x_* \equiv x_* - ky_*$.
  In the first line we have picked the fundamental domain in the covering space to be a parallelogram with the lower left corner at $(x_*,y_*)$.
 Here $g_{\cal C}^0$ is the transition function for $\phi_0$ along the cycle $\cal C$ and
\ie
n^0_{x\tilde x} ={1\over 2\pi } \left[ g^0_X(x+k ,y+1) - g^0_X(x,y)\right]= {1\over 2\pi }\left[ g^0_{\tilde X}(x+m,y)-g^0_{\tilde X}(x,y)\right]\,.
\fe
Recall that because of \eqref{transitioncon}, $\partial_xg_{X}^0$ (and similarly $\partial_x g_X$) is a function of one variable.

Alternatively, this operator can be written as
\ie
&{\cal U}(\phi_0)=
   \exp\Big[
{i\over 2\pi} \int_{y_*}^{y_*+1} dy \int_{\tilde x_*+ k y}^{\tilde x_*+ky+m} dx \, \phi_0(x,y) \partial_x\partial_y \phi(x,y) \\
&  - {i\over 2\pi}  \int_{\tilde x_* + ky_*} ^{\tilde x_* +ky_*+m} dx \, g_{\tilde X}^0(x, y_*) \partial_x \phi(x,y_* )\\
&+{i\over2\pi} \int_{y_*}^{y_*+1} ds  \,
\left[ g_X^0( \tilde x_* + ks  , s) k  \partial_x \phi ( \tilde x_* + ks , s)+\partial_y g_X^0( s  )  \,\phi ( \tilde x_* + ks  , s)\right]\Big]
\fe

In the special case of an untwisted torus, $k=0,m=1$, and this operator reduces to
\ie
{\cal U}(\phi_0)& = \exp\Big[
{i\over 2\pi} \int_{x_*}^{x_*+1} dx \int_{y_*}^{y_*+1} dy \, \phi_0(x,y) \partial_x\partial_y \phi(x,y) \\
&- {i\over 2\pi}  \int_{y_*}^{y_*+1}  dy  \, g_X^0 (x_*,y) \partial_y \phi(x_*,y)
- {i\over 2\pi}  \int_{x_*}^{x_*+1}  dx  \, g_Y^0 (x,y_*) \partial_x \phi(x,y_*)\\
&+ in_{xy}^0 \,\phi(x_*,y_*)
\Big]\,.
\fe

It is straightforward to check that this operator satisfies the following properties:
\begin{itemize}
\item It is independent of the reference point $(x_*,y_*)$.
\item It is symmetric under exchange of $\phi$ and $\phi_0$.
\item It is invariant under the gauge transformation of $\phi_0$:
\ie
&\phi_0 (x,y) \sim \phi_0(x,y) +2\pi n^x_0(x) +2\pi n^y_0(y)\,,\\
&n^x_0(x)\,, n^y_0(y)\in \bZ\,.
\fe
\item Since this operator is symmetric in $\phi\leftrightarrow \phi_0$, it is also invariant under the gauge transformation of $\phi$:
\ie
&\phi (x,y) \sim \phi(x,y) +2\pi n^x(x) +2\pi n^y(y)\,,\\
&n^x(x)\,, n^y(y)\in \bZ\,.
\fe
\item  If $\partial_x\partial_y \phi_0=0$, the operator ${\cal U}(\phi_0)$ depends only on the transition functions of $\phi$, where we have used the second property above. Therefore, it is a conserved operator in the $\phi$-theory.
\end{itemize}
When proving some of these statements, we drop integers of the form $\int dx g(x)\partial_x n(x)$ for some integer-valued functions $g(x),n(x)$ in the exponent of ${\cal U}(\phi_0)$.

We are now ready to discuss the most general conserved winding charges, which are ${\cal U}(\phi_0)$ with $\partial_x\partial_y\phi_0=0$.
As discussed in Section \ref{sec:special}, the most general such $\phi_0$ takes the form
\ie
\phi_0 ={2\pi r\over m} \left( \ThetaP (x,0,1) - k   \ThetaP(y,0,1)  \right)+f^x(x)+f^y(y)\,
\fe
with $r$ an integer modulo $m$ and $ f^i(x^i+1) = f^i(x^i)$.

The most general winding charge is therefore
\ie
{\cal U}(\phi_0) =&\exp\Big[
-{ ir\over m} \sum_{I=0}^{m-1}  g_{\tilde X}(I ,0)
+ {i r\over m}  \sum_{J=0}^{k-1} g_X( J,  0  )
\Big]\\
\times& \exp\left[{i} \oint dx f^x(x) Q^x(x) +  i \oint dy f^y(y)Q^y(y) \right]\,.
\fe
where we have set $x_*=y_*=0$ for simplicity.
We have thus unified the $U(1)$ charges \eqref{U1charges} and the  $\bZ_m$ charge \eqref{Zmcharge} into a single general winding operator ${\cal U}(\phi_0)$.

The analogous operators associated with the momentum symmetry of $\phi$ can be described using the dual field $\phi^{xy}$. (See \cite{paper1} for details on the self-duality of the $\phi$-theory.)   More explicitly, these operators are given by \eqref{Uphi0} with $\phi$ replaced by $\phi^{xy}$. They shift $\phi$ by $\phi_0$.

\section{Wilson operators of the $\bZ_N$ theory}\label{app:ZN}

In this appendix, we construct the most general gauge-invariant Wilson operator in the $\phi-A$ presentation of the $2+1$-dimensional $\bZ_N$ gauge theory on a twisted torus.  For simplicity, we will set $\lxeff=\lyeff=1$ in this appendix.

We follow a reasoning similar to that in Appendix \ref{app:Uphi0}.  We start with a background profile circle-valued function $\phi_0(x,y)$ and attempt to define
\ie
`` \exp\left[ {iN\over 2\pi} \int_{T^2} \phi_0(x,y) A_{xy}(x,y)\right]"
\fe
However, such an expression is generally not gauge-invariant, and depends on the choice of the fundamental domain of the torus.

To remedy these issues, we define the following operator in a similar spirit as in Appendix \ref{app:Uphi0}:
\ie
&{\cal U}(\phi_0)=
   \exp\Big[
{iN\over 2\pi}  \int_{y_*}^{y_*+1} dy \int_{\tilde x_*+ k y}^{\tilde x_*+ky+m} dx \, \phi_0(x,y) A_{xy} (x,y) \\
&  - {i\over 2\pi}  \int_{\tilde x_* + ky_*} ^{\tilde x_* +ky_*+m} dx \, g_{\tilde X}^0(x, y_*) \partial_x \phi(x,y_* )\\
&-{i\over2\pi}   \int_{y_*}^{y_*+1} ds  \,
\left[ g_X^0( \tilde x_* + ks  , s)   \partial_y \phi ( \tilde x_* + ks , s)+k\partial_x g_X^0( \tilde x_* + ks  , s)  \,\phi ( \tilde x_* + ks  , s)\right]\\
&+i  n^0_{x\tilde x} \, \phi(\tilde x_* + ky_* ,y_*) \Big]
\fe
When the equation of motion $A_{xy} = {1\over N}\partial_x\partial_y\phi  $ is imposed, it becomes the operator  ${\cal U}(\phi_0)$ in the $\phi$-theory \eqref{Uphi0}.
A similar calculation shows that ${\cal U}(\phi_0)$ is independent of the choice of the reference point $x_*,y_*$.
For simplicity, we will set $x_*=y_*=0$ from now on.

It is clearly invariant under $\phi\sim  \phi +2\pi n^x(x)+2\pi n^y(y)$.
Under a gauge transformation $\phi \sim \phi+N\alpha, A_{xy}\sim A_{xy} +\partial_x\partial_y \alpha$, this operator picks up a factor:
\ie
& \exp\Big[
{iN \over 2\pi}  \int_{y_*}^{y_*+1} dy \int_{\tilde x_*+ k y}^{\tilde x_*+ky+m} dx \, \alpha(x,y) \partial_x\partial_y \phi_0(x,y) \\
&  - {i N \over 2\pi}   \int_{\tilde x_* + ky_*} ^{\tilde x_* +ky_*+m} dx \, g_{\tilde X}^\alpha(x, y_*) \partial_x \phi_0(x,y_* )\\
&-{i N \over2\pi} \int_{y_*}^{y_*+1} ds  \,
\left[ g_X^\alpha( \tilde x_* + ks  , s)   \partial_y \phi_0 ( \tilde x_* + ks , s)+k\partial_x g_X^\alpha( \tilde x_* + ks  , s)  \,\phi_0 ( \tilde x_* + ks  , s)\right]\\
&+i  {N } n^\alpha_{x\tilde x} \, \phi_0(\tilde x_* + ky_* ,y_*) \Big]
\fe
where $g^\alpha_{\cal C}$ is the transition  function and    $n^\alpha_{x\tilde x}$ is similarly defined .
The condition for ${\cal U}(\phi_0)$ to be gauge invariant is
\ie\label{invphi0}
\partial_x\partial_y \phi_0=0\,,~~ { N  \phi_0 \over 2\pi}\in \bZ\,.
\fe
On top of these conditions, $\phi_0$ should still have $2\pi \bZ$-valued transition functions.
As discussed in Section \ref{sec:specialZN}, the most general such $\phi_0$ takes the form \eqref{generalZNphi}.

We now discuss several important examples of ${\cal U}(\phi_0)$:
\begin{itemize}
\item Consider
\ie
\phi_0 = {2\pi \over N } \left( \ThetaP(y,y_1,1) -\ThetaP(y,y_2,1)\right)\,.
\fe
(Here, for simplicity, we assume $0<y_2-y_1\le 1$.)
This gives the Wilson strip operator \eqref{stripoperatorZN}
\ie
\bW_{(y) }(y_1,y_2)= \exp\left[ i  \int_{y_1}^{y_2} dy \oint dx A_{xy}(x,y) \right] \,.
\fe
There is a similar choice of $\phi_0$ giving the Wilson strip that is extended along the $y$ cycle.
They generate the $\bZ_N$ dipole global symmetry of the $\bZ_N$ gauge theory.
\item As another example, we can take $\phi_0$ to be:
\ie
\phi_0  =  {2\pi \over \text{gcd}(N,m)} \left( \ThetaP(x,0,1)   - k \ThetaP(y, 0,1)   \right)\,.
\fe
This leads to the $\bZ_{\text{gcd}(N,m)}$ operator in \eqref{newAxy}:
\ie
{\bf U} =&
   \exp\Big[ \,
{i} { N\over   \text{gcd}(N,m)} \int_{0}^{1} dy \int_{ k y}^{ky+m} dx \,\left( \ThetaP(x,0,1)   - k \ThetaP(y, 0,1)  \right)  A_{xy} (x,y) \\
&-{i}{ m\over   \text{gcd}(N,m)} \int_{0}^{1} ds  \,
  \partial_y \phi (  ks , s) \,\Big] \,.
\fe
Note that the first integral has no support in the rectangle $k-1<x<k, 0<y<1$.
\end{itemize}

We have therefore unified the most general Wilson operators built from $A$ and $\phi$ into ${\cal U}(\phi_0)$ with different choices of $\phi_0$ obeying \eqref{invphi0}.
On a lattice, we have $L_x^{\rm eff}+L_y^{\rm eff}-1$ Wilson strip operators $\bW_{(i)}$ \eqref{stripoperatorZN} and one $\bZ_{\text{gcd}(N,M)}$ operator $\bf U$ \eqref{newAxy}.

 \section{Winding configurations of $\hat\phi$}\label{sec:hatwinding}

In this appendix,  we place classical circle-valued fields $\hat\phi^{i(jk)}$ in the $\mathbf{2}$ of $S_4$ on a twisted three-torus.
This, for example, is the quantum field of the $\hat\phi$-theory of \cite{paper2}.
In contrast to the parallel analysis for $\phi$ in Section \ref{sec:3+1winding}, we will see that, on a lattice, there is no new winding charge beyond those labeled by the $2(L_x^{\rm eff}+L_y^{\rm eff}+L_z)-3$ integer winding charges.

The $\hat\phi$-theory is dual to the $3+1$-dimensional $U(1)$ gauge theory of $A$ \cite{paper2}, where the winding charges of $\hat\phi$ are mapped to the electric charges of $A$.
Similar to the analysis of the transition functions in Section \ref{sec:2+1phiA}, there is no new electric charge in the $3+1$-dimensional gauge theory of $A$ beyond those labeled by the $2(L_x^{\rm eff}+L_y^{\rm eff}+L_z)-3$ integers.
Hence, the computation in this appendix provides another check of the above duality.

Furthermore, the gauge parameters $\hat \alpha^{i(jk)}$ of the gauge field $\hat A$ in \cite{paper2,paper3} are also in the $\mathbf{2}$ of $S_4$.
The winding configurations in this appendix were used as the large gauge transformations \eqref{hatgauge} in the gauge theory of $\hat A$ in Section \ref{sec:ZN}.

We now proceed to analyze the winding charges of $\hat\phi^{i(jk)}$.
The winding charges   obey $\partial_i \partial_j Q^k (x^i,x^j)=0$ with $i\neq j\neq k$ and can be expressed  as
\ie
&Q^x(y,z)= {1\over 2\pi}\oint dx \partial_x \hat\phi^{x(yz)}   = W^y_x(y) +W^z_x(z)\,,\\
&Q^y(z,x)= {1\over 2\pi}\oint dy \partial_y \hat\phi^{y(zx)}   = W^z_y(z) +W^x_y(x)\,,\\
&Q^z(x,y)={1\over 2\pi}\oint dz \partial_z \hat\phi^{z(xy)}   = W^x_z(x) +W^y_z(y)\,.
\fe
where  $W^i_j(x^i)\in \bZ$.
 Pairwise they share a common zero mode:
\ie\label{zeromode}
\left( \,W^i_k(x^i ) , W^j _k(x^j) \,\right)\sim \left( \,W^i_k(x^i )+1 , W^j _k(x^j)-1 \,\right)
\fe
Since $Q$'s are single-valued integer operators on the torus, the discussion in Section \ref{sec:special} implies that  $W^x_i(x+\lxeff) = W^x_i (x) ,~W^y_i(y+\lyeff) = W^y_i(y), ~W^z_i(z+\ell_z) = W^z_i(z)$.

However, these integers are not all independent. To see this, consider the following combination:
\ie
\partial_zW^z_y(z) - \langle \tilde X,Y\rangle \partial_zW^z_x(z)=& \partial_zQ^y  -\langle \tilde X,Y\rangle \partial_zQ^x \\
=& {1\over 2\pi } \partial_z \oint dy \partial_y \hat\phi^{y(zx)} - {\langle \tilde X,Y\rangle\over 2\pi} \partial_z\oint dx \partial_x\hat\phi^{x(yz)}\,.
\fe
We start by  showing that the last expression can be rewritten as
\ie\label{dzwind}
\hat Q^y(\hat Y)\equiv {1\over 2\pi } \partial_z \oint_{\cal C} \left( \partial_x \hat\phi^{y(zx)}  dx +\partial_y \hat\phi^{y(zx)} dy\right) \,,
\fe
where $\cal C $  is any curve homologous to $\hat Y\equiv m\tilde X=Y +\langle \tilde X,Y\rangle X$ \eqref{hatY}.
Note  that \eqref{dzwind} only depends on the homology class  of the curve $\cal C$, but not the explicit representative.
For any cycle $\cal S$, the charge $\hat Q^y({\cal S})$ computes the derivative of the winding charges of $\hat\phi^{y(zx)}$ along $\cal S$ and takes the form
\ie\label{generalhatQ}
\hat Q^y({\cal S}) = \sum_\gamma n_\gamma \deltaP(z, z_\gamma, \ell_z)\,,~~~~n_\gamma\in \bZ\,.
\fe

 Let us choose  $\cal C$ to be a curve that first goes around the $Y$ cycle once and then goes around the $X$ cycle $\langle\tilde X,Y\rangle $ times.
With this choice, we can write $\hat Q^y(\hat Y)$ as
\ie
\hat Q^y(\hat Y) &
= {1\over 2\pi } \oint_{\cal C}
\left(  \partial_z \partial_y \hat\phi^{y(zx)}  dy - \partial_x\partial_z \hat\phi^{x(yz)} dx - \partial_x  \partial_z\hat\phi^{z(xy)} dx\right) \\
&= {1\over 2\pi } \oint dy  \partial_z \partial_y \hat\phi^{y(zx)}   -
{\langle\tilde X,Y\rangle\over 2\pi}   \oint dx\partial_x\partial_z \hat\phi^{x(yz)}
\fe
In the first line we used $\hat\phi^{x(yz)} +\hat\phi^{y(zx)} +\hat\phi^{z(xy)}=0$.
In the second line we used the fact that $\partial_z\hat\phi^{z(xy)}$ is a single-valued operator, and therefore $\partial_x\partial_z\hat\phi^{z(xy)}dy$ does not contribute to the integral along $\cal C$, which is always aligned with the $x$ or the $y$ axes.  We have thus shown that
\ie
\partial_zW^z_y(z) - \langle \tilde X,Y\rangle \partial_zW^z_x(z)  = \hat Q^y(\hat Y) =  m \hat Q^y(\tilde X)\,.
\fe
Therefore, from the general form of the charge $\hat Q^y$ \eqref{generalhatQ}, we  arrive at the constraint:
\ie\label{hatWconstraint}
\partial_zW^z_y(z) - \langle \tilde X,Y\rangle \partial_zW^z_x(z)   =  m\sum_\gamma M_\gamma\deltaP(z, z_\gamma, \ell_z)\,,
\fe
for some integers $M_\gamma\in \bZ$.
Using \eqref{identity}, this constraint is equivalent to
\ie
\partial_zW^z_x(z) - \langle \tilde Y,X\rangle \partial_zW^z_y(z)   =  m\sum_\gamma M'_\gamma\deltaP(z, z_\gamma, \ell_z)\,,
\fe
for some integers $M'_\gamma\in \bZ$.

We summarize that the winding charges of $\hat\phi^{i(jk)}$ on a twisted torus are parameterized by $W^i_j(x^i)$ subject to the constraint \eqref{hatWconstraint} and the  redundancy \eqref{zeromode}.
On a lattice, we have $2(L_x^{\rm eff}+L_y^{\rm eff}+L_z)-3$ such integers.

Below we present the explicit winding configurations of $\hat\phi^{i(jk)}$ that realize all these charges.
For convenience, we will  use \eqref{zeromode} to gauge fix $W^z_x(z=0)=W^z_y(z=0)=W_z^x(x=0)=0$ below.

The most general winding configuration is
\ie\label{hatphiwinding}
&{1\over 2\pi} \hat\phi^{x(yz)}  = {\tilde y\over \ell_x}W^y_x(y)  + {x\over \ell_x}W^z_x(z) -  {y\over \ell_y}\, W^z_y(z) -  {z\over \ell_z}\, W^y_z(y),\\
&{1\over 2\pi}\hat\phi^{y(zx)}  =  {y\over \ell_y}W^z_y(z) + {\tilde x\over \ell_y} W^x_y(x)- {z\over\ell_z} \, W^x_z(x) -  {x\over\ell_x} \,W^z_x(z)\,,\\
&{1\over 2\pi}\hat\phi^{z(xy)}  = {z\over \ell_z} \left[ W^x_z(x) +W^y_z(y) \right]  -  {\tilde x\over \ell_y} \, W^x_y(x) -  { \tilde y\over \ell_x} \,W^y_x(y)\,.
\fe
where
\ie
&\tilde x = y- \langle \tilde Y,X\rangle {\ell_y\over \ell_x } x\,,\\
&\tilde y = x- \langle \tilde X,Y\rangle {\ell_x\over \ell_y} y\,.
\fe
These coordinates are shifted by $(\tilde x,\tilde y)\to (\tilde x -\langle\tilde Y,X\rangle \ell_y ,\tilde y+\ell_x)$ along the $X$ cycle, and are shifted by $(\tilde x,\tilde y) \to (\tilde x +\langle \tilde X,\tilde Y\rangle \ell_y , \tilde y)$ along the $\tilde X$ cycle.

Let us check that the transition functions are $2\pi \bZ$-valued.
The transition function around the $Z$ cycle is:
  \ie
&\hat  g^{x(yz)} _Z    =  -2\pi W^y_z(y)\,,\\
&\hat  g^{y(zx)} _Z   =  - 2\pi W^x_z(x)\,,\\
&\hat  g^{z(xy)} _Z    =  2\pi W^x_z(x) +2\pi W^y_z(y) \,.
  \fe
The transition function around the $X$ cycle is:
  \ie
&\hat  g^{x(yz)} _X     = 2\pi  W^y_x(y)  +2\pi   W^z_x(z) \\
&\hat  g^{y(zx)} _X     =-2\pi \langle\tilde Y,X\rangle W^x_y(x) -2\pi  W^z_x(z)    \\
&\hat  g^{z(xy)} _X   =   2\pi \langle \tilde Y,X\rangle W^x_y(x) - 2\pi W^y_x(y)\,.
  \fe
Finally, the  transition function  around the $\tilde X$ cycle is
  \ie
&\hat  g^{x(yz)} _{\tilde X}  =    2\pi  {\langle \tilde X,Y\rangle \over m}W^z_x(z) - 2\pi {1\over m} W^z_y(z)\\
&\hat  g^{y(zx)} _{\tilde X}   = 2\pi\langle\tilde X,\tilde Y\rangle   W^x_y(x) -2\pi {\langle\tilde X,Y\rangle \over m }W^z_x(z)  + 2\pi {1\over m}W^z_y(z) \\
&\hat  g^{z(xy)} _{\tilde X}    =   - 2\pi \langle\tilde X,\tilde Y\rangle W^x_y(x)
  \fe
  In the gauge choice $W^z_x(z=0)=W^z_y(z=0)=0$, \eqref{hatWconstraint} is equivalent to $ - \langle\tilde X,Y\rangle  W^z_x(z) + W^z_y(z)  \in m\bZ$. Hence, all these transition functions are indeed $2\pi\bZ$-valued.
The cocycle conditions are trivially satisfied since all the transition functions are single-valued.

\bibliographystyle{JHEP}
\bibliography{twisted}

\end{document}